\documentclass[]{jfm}

\usepackage{graphicx}
\usepackage{newtxtext}
\usepackage{newtxmath}
\usepackage{natbib}
\usepackage{hyperref}
\usepackage{subcaption}
\hypersetup{
    colorlinks = true,
    urlcolor   = blue,
    citecolor  = black,
}

\newcommand{\RomanNumeralCaps}[1]

\title{Response of a Turbulent Boundary Layer to a Synthetic Periodic Large-Scale Structure}

\author{Mitchell Lozier\aff{1}, Flint O. Thomas\aff{1} \and Stanislav Gordeyev\aff{1}\corresp{\email{sgordeye@nd.edu}}}
\affiliation{\aff{1}Department of Aerospace and Mechanical Engineering, University of Notre Dame, Indiana 46556, USA}

\begin{document}
\maketitle

% ABSTRACT =========================================
\begin{abstract}

The dynamic response of a zero-pressure gradient turbulent boundary layer (TBL) to a large-scale perturbation in the outer region was investigated experimentally. 
The baseline TBL had a moderate Reynolds number such that there was no naturally occurring energetic large-scale structure (LSS) present. 
An active plasma-based actuator was then placed in the outer region of the TBL to introduce a periodic, spanwise-uniform, synthetic LSS. 
This novel actuation scheme provides a new tool by which to experimentally examine the `top-down' view of TBL dynamics/interactions. 
The TBL response to this synthetic structure was investigated using a combination of planar particle imaging velocimetry and spanwise offset hot-wires, over a large streamwise extent downstream of the actuator device. 
Phase-locked analysis was implemented to isolate and measure the streamwise development of large-scale motions and changes in turbulence amplitude induced by this synthetic LSS. 
A strong correlation was observed between large-scale motions near the wall, linearly superimposed from the synthetic LSS, and a periodic modulation of turbulence amplitudes. 
This periodic modulation was found to be linked to phase-dependent changes in both the production and transport of turbulence driven by the induced large-scale motions. 
The phase speed of these induced large-scale motions, coupled with intermittent changes to spanwise coherence near the wall, revealed an additional, but transient, effect of the synthetic LSS on near-wall cycle dynamics. 
Overall, these results characterize the influences, and limitations, of top-down interactions on global TBL dynamics. 

\end{abstract}

% INTRODUCTION =========================================
\section{Introduction}
\label{sec:intro}

Wall-bounded turbulent flows, such as the turbulent boundary layer (TBL), play an important role in governing the performance of a number of engineered systems. 
For instance, TBL dynamics are directly coupled with relevant system properties such as skin friction drag, aero-optical distortion, noise generation, and heat transfer. 
Experiments conducted over many decades have established that quasi-periodic and repeating patterns of coherent vortical motions, broadly referred to as structures, exist within the TBL, as reviewed by \citet{robinson_coherent_1991}, and these structures are found to play a primary role in the production and maintenance of turbulence within the TBL. 
This stations the study of TBL structures as fundamental to understanding global TBL dynamics which, in turn, is prerequisite to optimizing strategies for their control and improving the performance of a wide range of engineered systems. 

Specifically, the dynamics of large-scale structures (LSS), and their effect on relevant TBL properties, has been investigated extensively \citep{adrian_vortex_2000, ganapathisubramani_investigation_2005, guala_large-scale_2006, Adrian_2007, hutchins_evidence_2007, Spencer2024}. 
The term LSS is typically used to describe the organized groups, or packets, of coherent vortices which are formed in wall-bounded flows at sufficiently high friction Reynolds numbers \citep[$Re_{\tau}>2000$,][]{robinson_coherent_1991, smits_highreynolds_2011, marusic_attached_2019}. 
Here, the term organized indicates that these coherent motions have characteristic length and time scales which can be measured. 
In the outer region of the TBL, the LSS can be identified by groups of spanwise-oriented vortices which form thin, inclined shear layers separating low- and high-speed regions of the flow, referred to as uniform momentum regions \citep{adrian_vortex_2000, de_silva_uniform_2016}. 
In high-Reynolds number TBLs, evidence of an organized and energetic LSS can also be seen in the premultiplied streamwise velocity energy spectra where a distinct peak emerges around the center of the log-linear region \citep[$y^{+}=3.9Re_{\tau}^{0.5}$,][]{Klewicki_2007, mathis_large-scale_2009} and over a range of relatively long streamwise wavelengths centered around $\lambda_{x}\approx4\delta$. 
In contrast, this outer peak is not present in the spectra of low-Reynolds number TBLs ($Re_{\tau}<2000$) which do not contain a naturally occurring energetic LSS as described above \citep{hutchins_evidence_2007}. 

In addition, there is a relatively smaller scale, organized and energetic structure, located closer to the wall, whose characteristics are largely independent of the Reynolds number \citep{hutchins_hot-wire_2009}. 
This near-wall coherent structure is characterized by unsteady, small-scale, quasi-streamwise vortices ($\lambda^{+}_{x}=1000$), which give rise to characteristic streaks of low- or high-speed fluid between them ($\Delta z^{+}=100$), contained within the buffer region \citep[$5<y^{+}<30$, ][]{kline_structure_1967, robinson_coherent_1991}. 
These near-wall vortices then create a significant spanwise variation in the streamwise velocity field (i.e., alternating low- and high-speed streaks) lending to their unsteadiness \citep{antonia_spanwise_1990}. 
These unsteady near-wall vortices also lead to intermittent and quasi-cyclic events whereby low-speed fluid is ejected away from the wall and high-speed fluid is transported towards the wall. 
These so-called burst-sweep events have been shown to be responsible for elevated levels of shear-stress in the near-wall region and for the significant production and maintenance of turbulence observed near the wall \citep{Luchik_1987, Chen_2021}. 
As such, the characteristics of these streamwise vortices and streaks, along with the dynamics of the bursting events, are intimately related to flow properties such as skin friction drag \citep{Kimfc_2011}. 

\begin{figure}
\begin{center}
\begin{subfigure}{0.36\textwidth}
    \centering
    \includegraphics[width=\textwidth]{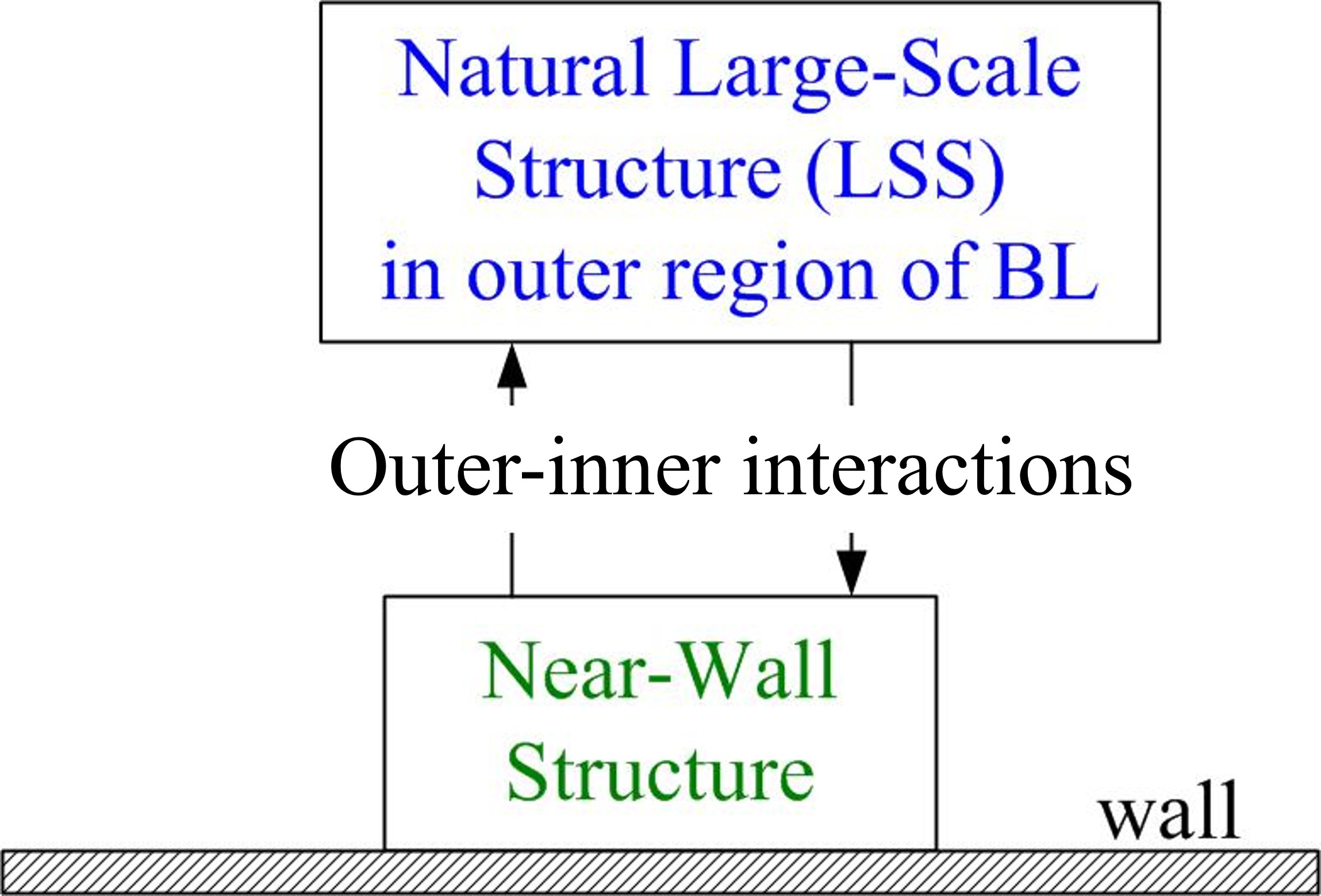}
    \caption{}
\end{subfigure}
\hspace{0.05\textwidth}
\begin{subfigure}{0.36\textwidth}
    \centering
    \includegraphics[width=\textwidth]{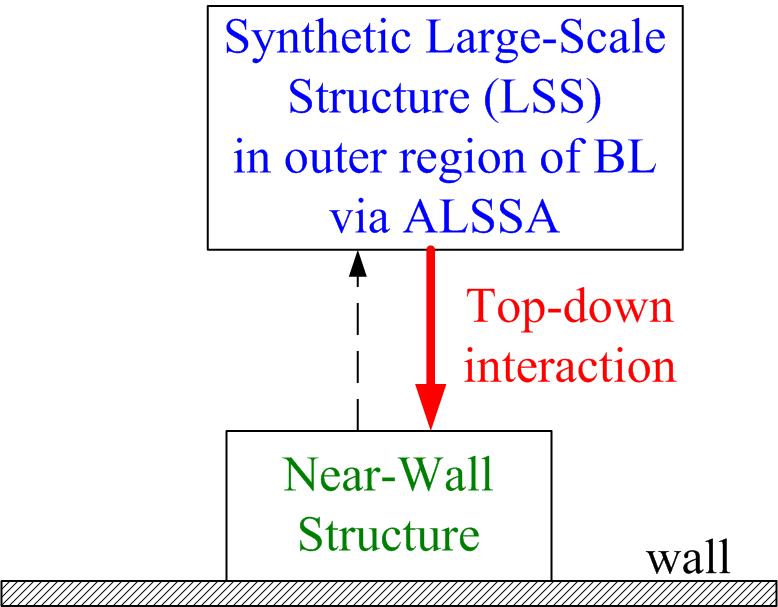}
    \caption{}
\end{subfigure}
\caption{(a) Schematic of outer-inner interactions between small-scale turbulent structures near the wall and the naturally-occurring LSS in the log-linear region of the TBL. (b) Schematic of the proposed approach to studying TBL dynamics via introducing a synthetic LSS.}
\label{fig:intro}
\end{center}
\end{figure}

Although several studies indicate that this near-wall turbulence production cycle is largely self-sustaining \citep{Hamilton_1995, jimenez_autonomous_1999, panton_overview_2001, Schoppa-Hussain_2002}, it has also been found that the LSS can influence near-wall turbulence dynamics through some outer-inner interaction mechanism \citep{adrian_vortex_2000}. 
This is shown schematically in figure~\ref{fig:intro}(a), whereby the distinct large- and small-scale structures of the TBL are coupled, to some extent, through outer-inner interactions \citep{robinson_coherent_1991, guala_large-scale_2006}. 
This conceptual model leads to two broad perceptions of global TBL structure. 
The first is a `bottom-up' view in which the near-wall mechanism for coherent structure generation is autonomous of the outer LSS dynamics, as supported by previous studies of the near-wall region \citep{panton_overview_2001, Schoppa-Hussain_2002, Thomas_et_al_2019}. 
The second is a `top-down' view whereby increasingly energetic large-scale motions/structures in the outer region govern near-wall turbulence dynamics through some top-down mechanism(s), the support for which is described in more detail in \citet{Hunt_2000} and \citet{corke_active_2018}. 
Underpinning this view, the dynamics of large- and small-scale motions within the log-linear and near-wall regions of the boundary layer have previously been shown, experimentally, to be correlated \citep{hutchins_large-scale_2007, mathis_predictive_2011}. 
Further, utilizing this observed correlation, a predictive model was developed in \cite{mathis_predictive_2011} to estimate near-wall streamwise turbulence statistics based on large-scale motions measured only within the log-linear region. 
The success of this model then suggests that there is a `universal' near-wall region which is modified (or modulated) through inter-scale interactions with these large-scale motions \citep[in the log-linear region and/or superimposed onto the near-wall region,][]{hutchins_large-scale_2007, mathis_predictive_2011, andreolli_separating_2023}.
This so-called modulation phenomena has also been measured experimentally \citep{mathis_large-scale_2009}, and near the wall there exists a distinct positive correlation between large-scale motions and the amplitude envelope of smaller-scale turbulent fluctuations. 
This correlation has also been found to extend to other regions of the TBL, other turbulent shear flows \citep{bandyopadhyay_coupling_1984}, and other flow properties such as the wall shear-stress \citep{mathis_estimating_2013}. 
The strength of this correlation was also found to increase with the Reynolds number \citep{mathis_large-scale_2009}, making these so-called classical amplitude modulating effects \citep[which represent a specific subset of the nonlinear inter-scale interactions coexisting within the TBL,][]{Lozier_AM_2024} increasingly important in technologically relevant flows. 
Overall, these seemingly competing views of TBL dynamics suggest that a clearer characterization of the mechanism(s) behind inter-scale outer-inner interactions is needed, particularly for the design and optimization of flow control strategies \citep{chernyshenko_quasi-steady_2012, corke_active_2018, andreolli_separating_2023}. 

To that end, active and passive flow control devices have also been used previously \emph{as tools} to study TBL dynamics by selectively perturbing the TBL structures and studying the resulting flow response. 
Some of the earliest passive devices used to modify the LSS in the TBL were large-eddy-breakup-devices (LEBUs) \citep{corke_new_1982, dowling_effect_1985, savill_manipulation_1988, kim_influence_2017, chan_large-scale_2022}. 
As reviewed by \citet{alfredsson_large-eddy_2018} and \citet{corke_active_2018}, these devices were typically made from thin plates, airfoils, or rods elongated in the spanwise direction and placed in the outer region of the boundary layer to disrupt large-scale motions and ultimately examine how near-wall dynamics were affected. 
Studies of these plate manipulators showed that they were effective (in certain conditions) at reducing near-wall turbulence intensity, reducing the frequency of bursting events, reducing local skin friction, and reducing streamwise boundary layer growth, among other effects \citep{corke_new_1982, dowling_effect_1985, savill_manipulation_1988, kim_influence_2017, chan_large-scale_2022}. 
In \cite{savill_manipulation_1988} it was shown that the interaction between the plate wake vortices and near-wall TBL structure was the primary mechanism for reducing skin friction drag; the maximum skin friction drag reduction always occurred close to the streamwise position where these wake vortices reached the sublayer. 

Recently, the effect of active flow control devices that alter/induce large-scale motions has also been studied \citep{mckeon_dynamic_2018, marusic_energy-efficient_2021, Meyers_2025}. 
For example, in a collection of experiments described in \citet{mckeon_dynamic_2018}, a vertically oscillating element, referred to as a dynamic roughness element, was used to introduce a controlled traveling wave into the near-wall and log-linear regions of the TBL. 
Measurements of the synthetic modes highlight the periodic nature of these synthetic large-scale motions, introduced via the dynamic roughness element, which allowed for systematic study of their interactions with small-scale turbulence. 
Measurements of the correlation between these modes and smaller-scale turbulent motions showed that the synthetic large-scale motions had a strong effect on triadically coupled small-scale motions, and a phase-locking effect was observed in the near-wall structure of the TBL as a result \citep{duvvuri_triadic_2015}. 
Alternatively, motivated by intervention in the streak transient growth autonomous near-wall turbulence production mechanism proposed by \cite{Schoppa-Hussain_2002}, pulsed-DC plasma actuation has also been used for active flow control, producing spanwise blowing confined to the buffer region of the TBL as described in \citep{Thomas_et_al_2019}. 
This approach has resulted in up to 78\% viscous drag reduction, and measurements of the mechanism for the observed drag reduction presented in \cite{Duong_2021} and \citet{Meyers_2025} were consistent with intervention in the streak transient growth mechanism. 
Scaling laws developed from this method were also verified in subsequent airfoil experiments at freestream Mach numbers up to $M_{\infty} = 0.5$ \citep[as reported in][]{Thomas_et_al_2023}, and a similar plasma actuation approach utilized by \cite{Cheng_et_al_2021} to introduce near-wall streamwise vorticity also yielded significant viscous drag reduction in a flat plate TBL. 
Collectively, the results of these active flow control experiments (plasma actuation and dynamic roughness) demonstrate the significant effects that induced large-scale perturbations in the near-wall and log-linear regions can have on near-wall turbulence characteristics. 

In \cite{ranade_turbulence_2019} it was observed that large-scale perturbations imposed from \emph{outside} of the boundary layer could also modify near-wall turbulence characteristics. 
In this particular study, the TBL was `forced' via the unsteady pressure field generated by a forced shear layer located outside of the boundary layer ($y\approx4\delta$). 
The turbulence intensity inside the boundary layer was found to be both amplified and modulated by the this external forcing. 
This result demonstrates that, like the internal (near-wall) perturbations/forcings described above, large-scale perturbations to the outer region of the TBL can also affect the amplitude and organization of structures within the rest of the TBL, and importantly in the near-wall region. 

Inspired by these studies, an active flow control device in the form of a plasma-based actuator was used in the current study to introduce synthetic, periodic, spanwise-uniform large-scale motions into the outer region of the TBL to investigate the mechanism of outer-inner interaction. 
It is important to note that the baseline TBL examined in this study had a Reynolds number which was purposely low enough that there were no \emph{naturally occurring} energetic LSS in the log-linear region. 
Instead, the plasma actuator device was designed to introduce a controlled, spanwise-uniform perturbation, with variable streamwise wavelength (i.e., frequency) and amplitude, into the outer region of the boundary layer at a user selected wall-normal location. 
In this way, the top-down portion of the outer-inner interaction can be isolated and studied directly as modeled in figure~\ref{fig:intro}(b). 
This approach was tested previously, and the effect of this actuator device on the baseline canonical TBL was characterized parametrically, in \cite{lozier_experimental_2023, lozier_response_2024}. 
These preliminary results confirmed that the actuator device could produce spanwise-uniform large-scale motions, with a range of user selected wavelengths, and the introduction of these large-scale motions in the outer region resulted in modulation of the turbulence intensity (i.e., periodic changes in the magnitude of the turbulence intensity were observed) across the TBL, including the near-wall region, consistent with the other active flow control approaches described above. 
However, in contrast to previous approaches, the present approach enables the production of a perturbation which is more closely related to the naturally occurring LSS found in higher Reynolds number TBLs. 
Specifically, the actuator design and methodology allow for the generation of spanwise vortices, confined to the outer region, with streamwise wavelengths which fall within the broad hierarchy of canonical large-scale motions, without introducing any direct/mechanical disruptions to the near-wall region. 
It is noted that other characteristics of the canonical LSS, such as their inherent three-dimensionality and stochastic spatial/temporal distribution (i.e., hairpin-like vortices, grouping/packets, meandering behaviors, etc.) can not be replicated with the present actuator design, which is why a spanwise-uniform synthetic LSS is preferred instead. 
In spite of these limitations, the synthetic perturbation produced here is well suited for investigating outer-inner interactions, in general, and the contributions of certain canonical large-scale motions to global TBL dynamics (i.e., from a specific subset of the full hierarchy). 
An especially significant advantage to the present approach is that the imposed perturbation is periodic which provides a convenient phase reference for separating the induced large-scale motions and the TBL's response to them. 
To that end, previous experimental \citep{lozier_response_2024} and numerical \citep{liu_spatial_2022} studies have demonstrated that for a range of actuation frequencies and wall-normal forcing locations, significant modulation of the near-wall turbulence intensity can be achieved, phase-locked with the imposed perturbation in the outer region, confirming the top-down nature of these modulation effects (i.e., the modulation phenomena near the wall was found to be exclusively governed by the characteristics of the synthetic LSS which were confined to the outer region). 
Motivated by these results, the present study aims to now investigate the process(es) by which these well-defined large-scale vortical motions in the outer region both superimpose onto, and modulate turbulence intensities within, the near-wall and log-linear regions. 
This study further aims to investigate if these imposed large-scale motions significantly influence the characteristics/dynamics of the canonical autonomous near-wall structure/cycle. 
Because this actuation scheme targets only the top-down type interactions, observations regarding the effect of the synthetic LSS on the near-wall turbulence characteristics/dynamics will ultimately test the extent to which LSS can abate, reorganize, or otherwise influence the near-wall turbulence dynamics, particularly in the context of optimizing flow control strategies.

% EXPERIMENTAL SETUP =========================================
\section{Experimental Setup}
\label{sec:setup}

\begin{figure}
\centering
\begin{subfigure}{\textwidth}
    \centering
    \includegraphics[width=0.7\textwidth]{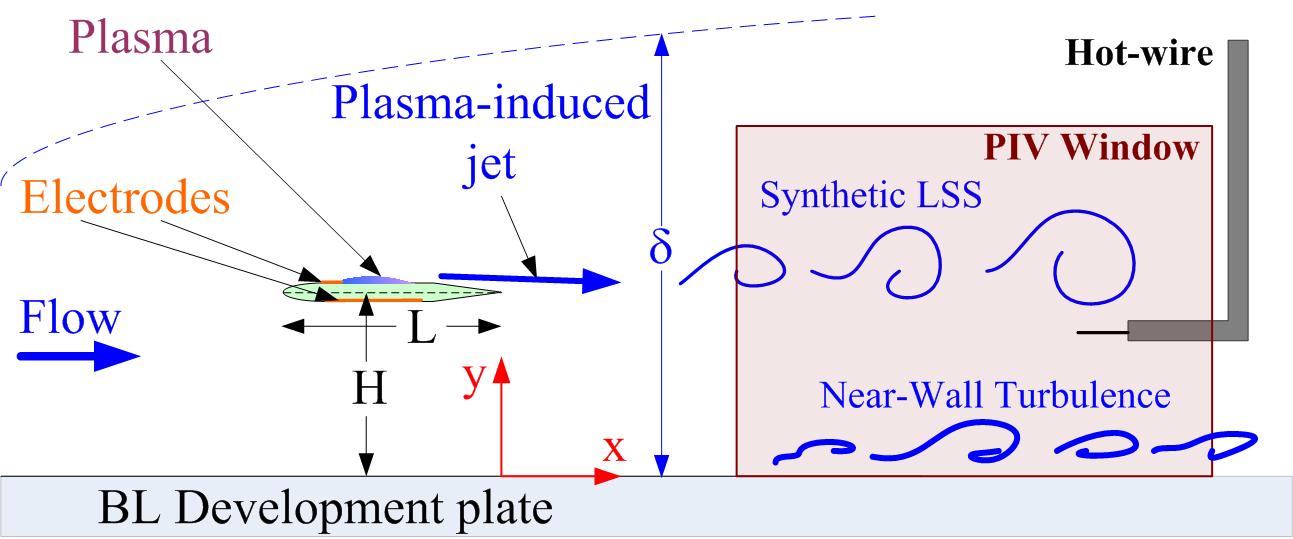}
    \caption{}
\end{subfigure}
\begin{subfigure}{0.3\textwidth}
    \centering
    \includegraphics[width=\textwidth]{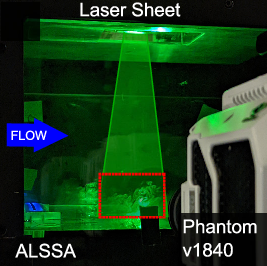}
    \caption{}
\end{subfigure}
\hspace{0.08\textwidth}
\begin{subfigure}{0.3\textwidth}
    \centering
    \includegraphics[width=\textwidth]{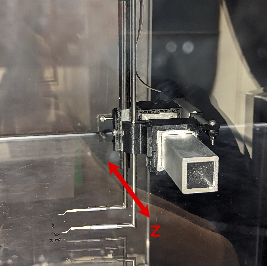}
    \caption{}
\end{subfigure}
\caption{(a) Schematic of experimental setup. (b) Photograph of PIV setup. (c) Photograph of spanwise offset hot-wire setup.}
\label{fig:setup}
\end{figure}

The experiments presented in this study were performed in one of the low-turbulence, subsonic, in-draft wind tunnels located at the Hessert Laboratory for Aerospace Research at the University of Notre Dame. 
Experiments were performed in a test section of $0.6$ m square cross-section and $1.82$ m length with nominal free-stream turbulence levels of less than $0.1\%$. 
The experimental setup is shown schematically in figure~\ref{fig:setup}(a). 
For this study, a flat, 2-meter-long, boundary layer development plate, which spanned the full test section width, was installed at the mid-height of the test section. 
The boundary layer development plate had a leading edge which was rounded into a semicircular profile with sandpaper distributed roughness attached in order to facilitate rapid turbulent transition on the top side. 

A plasma-based active flow control actuator, as described below, was attached to the top side of the boundary layer development plate at a fixed streamwise location of $140$ cm downstream of the leading edge of the boundary layer development plate. 
This location was chosen to ensure both a long upstream TBL development length and a sufficient streamwise fetch for measurements before end of the boundary layer development plate. 
The freestream velocity was also measured to be constant between the actuator location and the end of the boundary layer development plate confirming baseline zero-pressure gradient (ZPG) conditions throughout the measurement region. 

A set of representative baseline turbulent boundary layer parameters were measured at $x=3\delta$ downstream of the actuator trailing edge ($150$ cm downstream of the leading edge of the development plate) using a single hot-wire probe \citep[additional details given in ][]{lozier_response_2024}. 
These parameters are summarized in table~\ref{tbl:params}, for reference. 
The skin friction velocity $u_{\tau}$ and friction coefficient $c_{f}$ were determined using the Clauser method. 
The boundary layer thickness, $\delta$, is taken as the classical 99\% thickness here. 
In all the experiments to be described in this study, the wind tunnel inlet free-stream velocity was set to $U_\infty = 7$ m/s and was measured to be within $\pm 1\%$ of the expected free-stream velocity before each measurement session. 

\begin{table}
  \begin{center}
    \caption{Representative canonical turbulent boundary layer parameters at $x=3\delta$.}
    \label{tbl:params}
    \begin{tabular}{ccccccccc}
      \hline
      $\delta$ & $\delta^{*}$ & $\theta$ & $U_{\infty}$ & $u_{\tau}$ & $c_{f}$ & $H_{S}$ & $Re_{\theta}$ & $Re_{\tau}$ \\ 
      $34$ mm & $5.3$ mm & $3.9$ mm & $7$ m/s & $0.31$ m/s & $0.0039$ & $1.36$ & $1770$ & $690$ \\ 
      \hline
    \end{tabular}
  \end{center}
\end{table}

\begin{figure}
\centering
\begin{subfigure}[c]{0.45\textwidth}
    \centering
    \includegraphics[width=0.6\textwidth]{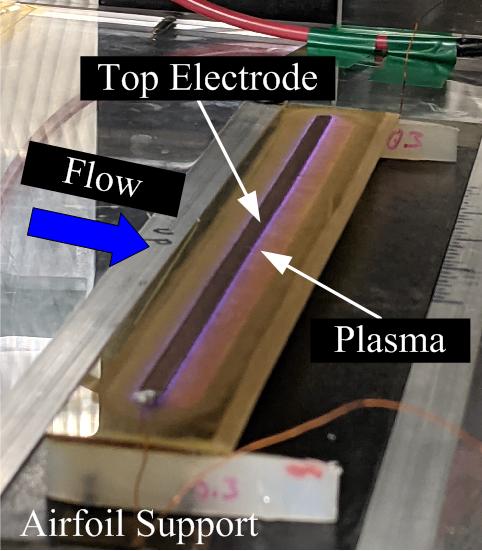}
    \caption{}
\end{subfigure}
\begin{subfigure}[c]{0.45\textwidth}
    \centering
    \includegraphics[width=\textwidth]{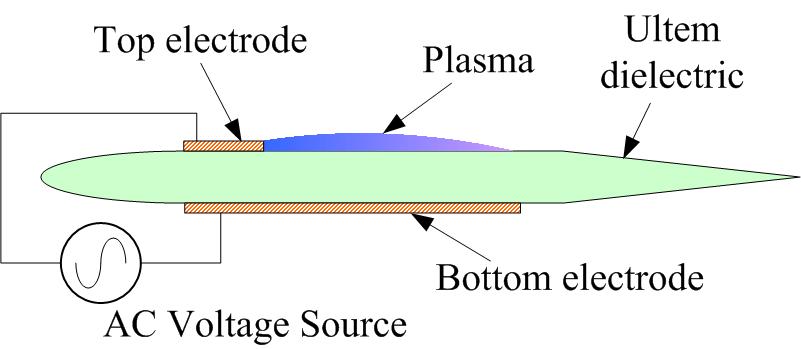}
    \caption{}
\end{subfigure}
\caption{(a) Photograph of ALSSA. (b) Schematic of an AC-DBD plasma actuator showing essential components (electrode thicknesses are exaggerated for clarity).}
\label{fig:alssa}
\end{figure}

As shown in figure~\ref{fig:alssa}, the plasma-based Active Large-Scale Structure Actuator (ALSSA) was used in this study to introduce periodic, spanwise-uniform plasma–induced forcing into the outer region of the TBL. 
The actuator plate was made from a $2$ mm thick sheet of Ultem, dielectric polymer. 
The dimensions of the actuator plate were $W=25$ cm ($W=8\delta$, $W^{+}=5600$) width in the spanwise direction and $L=32$ mm ($L\approx 1\delta$, $L^{+}=640$) length in the streamwise direction. 
This actuator width is sufficient to conservatively ensure that an approximately $5\delta$ wide region of spanwise uniform flow is maintained across all downstream measurement locations, despite the three-dimensional end effects from the finite length actuator plate as demonstrated in \cite{lozier_response_2024}. 
The streamwise extent and thickness of the actuator plate were minimized, within safe margins for plasma generation, the leading edge of the dielectric plate was rounded, and the last $10$ mm of the trailing edge were linearly tapered to a half-angle of $10\deg$ to minimize the wake downstream of the actuator plate. 
The plasma actuator was supported above the boundary layer development plate by twin NACA0010 airfoils which were fabricated at user-selected heights such that the distance between the actuator plate and wall was fixed at $H=0.3\delta$ ($H^{+}=200$) for these experiments.
This wall-normal position of the ALSSA device was found to induce the strongest response of the TBL in the near-wall region which was deemed optimal for the aims of the current study. 
Additional details about the optimization of the ALSSA device can be found in \cite{lozier_response_2024}. 

Electrodes were attached to the dielectric plate in order to generate a spanwise-uniform plasma jet on the top side of the actuator as shown schematically in figure~\ref{fig:alssa}(b). 
An upper surface electrode of $0.05$ mm thick copper foil tape was placed $15$ mm from the dielectric plate leading edge and was $4$ mm in length (streamwise) and $22$ cm in width (spanwise). 
On the bottom surface, a second copper foil electrode was aligned with the top electrode but was $12$ mm in length. 
Additional details regarding the construction and operation of the actuator can be found in \cite{lozier_characterization_2023}.  

Alternating current dielectric barrier discharge (AC-DBD) plasma was produced by connecting the copper electrodes on the top and bottom of the actuator to a high voltage AC source \citep{thomas_optimization_2009}, as shown in figure~\ref{fig:alssa}(b), which provided a $40$ kV peak-to-peak sinusoidal waveform excitation to the electrodes at a frequency of $4$ kHz. 
The peak-to-peak voltage from the high voltage AC system was measured during experiments and maintained within $\pm 5\%$ of the expected excitation voltage. 
As shown schematically in figure~\ref{fig:setup}(b), a plasma jet was formed above the top surface of the actuator plate above the exposed portion of the lower surface electrode. 
At the $4$ kHz actuation carrier frequency, the plasma actuator operates in a quasi-steady mode, essentially creating a spanwise-uniform, steady, plasma-induced jet above the actuator plate directed in the streamwise direction \citep{thomas_optimization_2009}. 
To introduce the desired periodic forcing, the sinusoidal waveform used to generate the plasma jet was modulated by a square wave with a fifty percent duty cycle oscillating at the desired forcing frequency. 
In the experiments presented here, the forcing frequency was fixed at $f_{P}=80$ Hz ($f_{P}\delta/U_{\infty}=0.39$), which was again found to create the strongest response of the TBL within the near-wall region which is optimal for this study. 
A detailed discussion for this choice is provided in \cite{lozier_response_2024}. 

Because the physical presence of the actuator plate alone may alter characteristics of the canonical TBL, three measurements of the boundary layer were taken to systematically analyze the effect of the plasma actuation. 
The first measurement was of the canonical TBL (no actuator plate) which served as a baseline. 
Next, measurements were taken with the actuator plate installed, but no active plasma forcing, which will be referred to as the `plate only' case. 
Finally, measurements were taken with active plasma forcing, under the conditions described above, which will be referred to as the `plasma on' case. 

\subsection{Particle Imaging Velocimetry}

In all particle image velocimetry (PIV) experiments reported here, a double-pulsed $532$ nm laser (EverGreen 200 laser with $200$ mJ pulse and $10$ ns pulse width) was used to make phase-locked, spatially resolved, planar flow field measurements. 
The flow was seeded with DEHS particles (diameter $<1\;\mu \mathrm{m}$) through the tunnel inlet which were illuminated in the streamwise wall-normal plane. 
Diverging laser sheets (half angle of $10^\circ$) were formed and directed through the top of the test section and down onto the boundary layer development plate resulting in a spanwise thickness of $<2$ mm at the boundary layer development plate, see figure~\ref{fig:setup}(b). 
The overlap of these laser sheets resulted in a maximum measurement region of $x=60\times y=24$ mm ($x=1.8\times y=0.7\delta$). 
Seven measurement regions along the same wall-normal/streamwise plane were used with a minimum overlap of 10 percent ($\Delta x_{overlap} \geq 0.2\delta$) between neighboring regions, resulting in a total experimental streamwise measurement region between approximately $x=0.2 - 10\delta$ downstream of the actuator (with the leading edges of these regions located at approximately $x/\delta$ = -0.1, 1.1, 2.5, 4.0, 5.5, 6.8 and 8). 
The closest reliable measurement point to the actuator was limited to $x\approx0.2\delta$ downstream of the trailing edge of the actuator due to optical constraints of the experimental setup. 

The separation between consecutive laser pulses was $\Delta t=50\mu \mathrm{s}$ and laser pulses were created with a repetition frequency of $12.5$ Hz, phase-locked with the plasma forcing cycle. 
Images were captured with a Phantom v1840 high-speed camera at a sampling frequency of $100$ Hz (phase-locked with both the laser pulses and plasma forcing cycle) and with a resolution of $1920\times 1080\;\mathrm{px}^{2}$.
This resulted in a final measurement resolution of $12\Delta x^{+}\times12\Delta y^{+}$.
Additional details of the camera, lens, and processing parameters used in these experiments can be found in \cite{lozier_characterization_2023}. 
The error of the computed velocity vectors was estimated to be $\epsilon_{u}/U_{\infty}<0.5\%$ ($\epsilon_{u}<0.1u_{\tau}$). 
As mentioned above, the PIV image acquisition and plasma forcing cycles were phase locked, and due to the specific frequency chosen for the laser pulse repetition, there were five distinct phases of the plasma forcing cycle ($\phi=0,2\pi /5,4\pi /5,6\pi /5$ and $8\pi/5$ radians) measured non-sequentially over the course of 32 actuation periods. 
Over the total sampling time of $T_{s}=480$ s, each phase of the plasma forcing cycle was measured $n=1200$ times, to obtain reasonable convergence of the phase-locked statistics. 
This PIV configuration resulted in measurements of the two-dimensional velocity field that were phase-locked, spatially resolved in the streamwise and wall-normal directions, and measured over a long streamwise extent downstream of the actuator. 

\subsection{Hot-Wire Anemometry}

To measure spanwise spatial correlations of the streamwise velocity, a constant temperature anemometer (CTA) with two boundary layer style hot-wire probes (Dantec 55P15) with diameters of $5\;\mu \mathrm{m}$ and lengths of $l=1.5$ mm ($l^{+}=30$) were used to collect time series of the streamwise component of velocity across a range of spanwise separations, see figure~\ref{fig:setup}(c). 
The sampling frequency was $f_{S}=30$ kHz, with a low-pass filter set to 10 kHz, and the sampling time was $T_{S}=90$ s for each case such that $t^{+}=u_{\tau}^{2}/(f_{S}\nu) \approx 0.2$ and $T_{S}U_{\infty}/\delta \approx 19000$. 
The first hot-wire was held stationary at the mid span of the test section (which also corresponds to mid span of the actuator). 
The second hot-wire, mounted on a spanwise oriented traversing stage, was attached to the stationary hot-wire and matched in x- and y-position. 
Measurements were taken simultaneously with the hot-wires starting at a minimum separation of $\Delta z=2$ mm ($\Delta z^{+}=40$) and reaching a maximum separation around $\Delta z=1.1\delta$ ($\Delta z^{+}=760$). 
For all of the results presented here the probes were positioned at the farthest downstream location achievable, $x=8\delta$, as measured downstream of the plasma actuator’s trailing edge.

% DATA REDUCTION =========================================
\section{Data Reduction}
\label{sec:data}

Phase-locked analysis was performed independently at each $(x,y,z)$ location, for both the PIV and hot-wire measurements. 
Since the actuator introduces periodic forcing into the flow, it is both convenient and necessary to phase-lock measurements to the periodic forcing frequency in order to separate the induced large-scale motions from the response of the TBL \citep[across a broad hierarchy of scales;][]{ranade_turbulence_2019}. 
To implement phase-locked analysis, a triple Reynolds decomposition of the streamwise velocity was utilized,
\begin{equation}
    u(y,t)=U(y)+\tilde{u}(y,\phi)+u'(y,\phi,n),
    \label{eq:decomp}
\end{equation}
\noindent where $u$ is the instantaneous streamwise velocity, $U$ is the time mean component of velocity, $\Tilde{u}$ is a phase dependent or modal velocity component, $u'$ is a velocity fluctuation, and $\phi$ is the phase of the plasma actuation cycle. 
Here $\phi$ can be defined by
\begin{equation}
    \phi=\left(\frac{t_{n}}{T_{P}}-n\right)2\pi,
    \label{eq:phase}
\end{equation}
\noindent where $n$ represents an actuator period ($T_{p}=1/f_{p}$) length sample, or realization, of the original velocity time series. 
These $n$ realizations are extracted from the full time series such that the beginning of each realization is aligned with the plasma being turned on in the actuation cycle. 
In \eqref{eq:phase}, $t_{n}$ is a time in the $n^{th}$ realization, which is related to the phase angle, $\phi$, by the period of the actuation cycle. 
$n$ realizations of the mean removed velocity can be ensemble averaged to find the modal component of velocity as a function of the phase angle,
\begin{equation}
    \tilde{u}(y,\phi)=\langle u(y,\phi,n)-U(y)\rangle_{n}.
    \label{eq:modal}
\end{equation}
\noindent Here the angle brackets denote ensemble averaging over all realizations. 
This modal velocity will be used to quantify the periodic, large-scale perturbation/motions induced by the actuator. 
The velocity fluctuations, $u'$, that remain after removing both the modal and mean component of velocity from each realization were then used to compute an ensemble-averaged root-mean-square (RMS),
\begin{equation}
    u'_{RMS}(y,\phi)=\sqrt{\langle u'(y,\phi,n)^{2}\rangle_{n}}.
    \label{eq:residual}
\end{equation}
\noindent We will refer to this quantity as the residual turbulence amplitude. 
The phase-averaged mean was also removed from the residual turbulence amplitude to define a phase-dependent \emph{change} in the residual turbulence amplitude,
\begin{equation}
    \Delta u'_{RMS}(y,\phi)=u'_{RMS}(y,\phi)-\langle u'_{RMS}(y,\phi)\rangle_{\phi}.
    \label{eq:residual change}
\end{equation}
\noindent This new quantity is especially useful because it reveals changes in the distribution of velocity fluctuations that are phase-locked with the plasma actuation cycle. 
Further, removing the phase-averaged mean from the residual turbulence amplitude makes it easy to compare relative changes in the turbulence intensity across all wall-normal locations. 
As such, this new quantity \eqref{eq:residual change}, referred to concisely as the residual turbulence, will be used to quantify the TBL's response to the periodic perturbation induced by the actuator. 
The phase-locking treatments presented thus far will also be applied to other flow field quantities such as the wall-normal velocity and Reynolds stress in later analysis; \eqref{eq:modal} and \eqref{eq:residual}, presented above, can be used to analyze any quantity of interest but have been shown with the streamwise component of velocity here for reference. 

It should be noted here that velocity fluctuations can alternatively be decomposed using methods such as scale- (or frequency-) based filtering \citep{mckeon_dynamic_2018}, to separate the motions which are induced by the plasma forcing from the smaller-scale turbulent motions of the TBL. 
While employing a frequency-based filtering method is not feasible for the present PIV data (insufficient temporal resolution) this method was tested extensively with the hot-wire data from previous experiments. 
The results from frequency-based filtering and the phase-locked analysis presented here were compared and identical behaviors/results were observed for all streamwise and wall-normal locations, except for those in the immediate vicinity of the actuator plate \citep{lozier_characterization_2023}. 
The discrepancies in the immediate vicinity of the actuator plate likely stem from the small-scale turbulence which is directly introduced by the actuator plate's wake. 
However, the results presented in the following section are primarily focused on the region between the actuator plate and the wall. 
Additionally, the residual turbulence amplitude was found to contain a relatively small contribution from cycle-to-cycle variability in the synthetic LSS, in addition to the primary contribution from smaller-scale turbulent motions. 
However, this contribution is \emph{not phase-dependent} and therefore removed from the residual turbulence in \eqref{eq:residual change}. 
As such, the modal velocity and residual turbulence which are obtained from this phase-locked analysis of the PIV (and hot-wire) data are expected to be (approximately) equivalent to the results which would be obtained from a frequency-based filtering method. 
Further, this agreement supports the interpretation that the new residual turbulence quantity is representative of the TBL's response (specifically from small-scale turbulent motions) to the periodic perturbation induced by the actuator.

% RESULTS =========================================
\section{Results}
\label{sec:result}

\begin{figure}
  \begin{center}
    \includegraphics[width=1\textwidth]{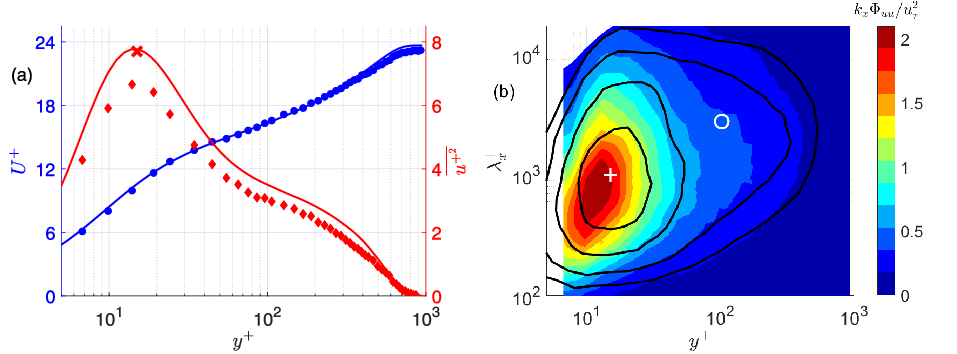}
    \caption{(a) Streamwise mean velocity and turbulence intensity for the baseline canonical TBL \citep{lozier_response_2024}. Solid lines are $Re_{\tau}=690$ DNS data from \citet{jimenez_turbulent_2010}. Red `$\times$' is peak turbulence intensity corrected for hot-wire spatial filtering effects \citep{hutchins_hot-wire_2009}. (b) 1-D premultiplied energy spectra \citep{lozier_response_2024}. Plus marks inner peak ($y^{+}=15, \lambda_{x}^{+}=1000$). Open circle marks theoretical outer peak location ($y^{+}=3.9Re_{\tau}^{0.5}, \lambda_{x}^{+}\approx 2700$). Contour lines are $Re_{\tau}=1100$ experimental data from \citet{hutchins_large-scale_2007}. $x=3\delta$.}
    \label{fig:canonical}
  \end{center}
\end{figure}

The baseline canonical TBL for this experiment was quantified using a single hot-wire probe in \citet{lozier_response_2024}, and a statistical and spectral representation of this baseline TBL are shown in figure~\ref{fig:canonical}, for reference. 

The inner-variable scaled streamwise mean velocity and turbulence intensity profiles in figure~\ref{fig:canonical}(a) match the DNS results from \citet{jimenez_turbulent_2010} at a comparable friction Reynolds number when the effect of the finite hot-wire length is accounted for (note the red `$\times$') as recommended by \citet{hutchins_hot-wire_2009}. 
The extent of the log-linear region for the baseline TBL was determined to be between $50 < y^{+} < 200$ and the mean geometric location was determined to be at $y^{+}=100$ which is very close to the expected value of $y^{+}=3.9Re_{\tau}^{0.5}$ \citep{mathis_large-scale_2009, Klewicki_2007}. 
Similarly, the premultiplied 1-D wavenumber energy spectra in figure~\ref{fig:canonical}(b) agrees well with experimental data from \citet{hutchins_large-scale_2007} with a similar Reynolds number. 
Most importantly, in figure~\ref{fig:canonical}(b), is the lack of a naturally occurring outer LSS energy peak for the baseline TBL, as desired for this study. 

\begin{figure}
  \begin{center}
    \includegraphics[width=1\textwidth]{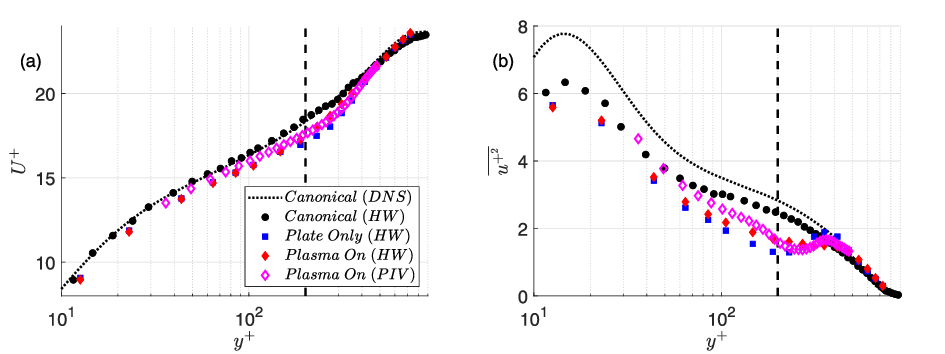}
    \caption{Comparison of canonical, plate only, and plasma on statistics measured using hot-wire and PIV. (a) Streamwise mean velocity and (b) turbulence intensity. Hot-wire data from \cite{lozier_response_2024}. Matched $Re_{\tau}$ DNS from \cite{jimenez_turbulent_2010}. The vertical dashed line represents the location of the actuator plate. $H^{+}=200,\ f_{p}=80$ Hz, $x=5\delta$.}
    \label{fig:stats}
  \end{center}
\end{figure}

A statistical description of the effects of the ALSSA device on this baseline canonical TBL, from single-point hot-wire measurements, has been given in \citet{lozier_response_2024}. 
The streamwise mean velocity and turbulence intensity from \citet{lozier_response_2024} for the canonical, plate only, and plasma on cases are replotted here in figure~\ref{fig:stats}. 
When the actuator plate is added to the baseline TBL, with no active forcing, there is a decrease in both the mean velocity and turbulence intensity around the actuator plate location (vertical dashed line in figure~\ref{fig:stats}) consistent with a low momentum deficit wake being induced by the actuator plate \citep[also consistent with previous studies of LEBU devices,][]{corke_new_1982, chan_large-scale_2022}. 
When active plasma forcing is added, there is some recovery in both the streamwise mean velocity and turbulence intensity, towards the canonical levels, especially above the plate location ($y^{+}=200$) where the plasma jet is located ($y^{+} \approx 265$). 
It is also noted that, near the wall ($y^{+}<40$), there is no effect of the plasma forcing on the (time-averaged) turbulence intensity, as compared to the plate only case, at this streamwise location; however, in previous studies, an increase in the peak plasma on turbulence intensity was observed farther downstream \citep[e.g., at $x=8\delta$ where an increase was also observed for the time-averaged Reynolds stress and turbulence production;][]{lozier_characterization_2023}. 

Next, results from the planar PIV measurements of the region downstream of the actuator will be presented to both complement and expand upon these earlier results. 
These new PIV measurements are unique from the previous measurements in a few key ways. 
First, they are spatially resolved in the streamwise direction, over a large streamwise extent, meaning the true spatially evolving response of the TBL can be discerned. 
They also remain phase resolved, although at a much lower resolution than previous measurements. 
And finally, direct measurement of the wall-normal component of the velocity (in addition to the streamwise component), over a spatially resolved two dimensional field allows for direct calculation of new flow properties such as the Reynolds shear stress and vorticity. 

Statistical profiles from the plasma on case, measured using the planar PIV system, are validated here by comparison with the results from \citet{lozier_response_2024} in figure~\ref{fig:stats}. 
The streamwise mean velocity at a single downstream location is shown in figure~\ref{fig:stats}(a) and there is good agreement between the plasma on profile measured using PIV and the earlier measurement using a single hot-wire. 
Similarly, there is good agreement in the streamwise turbulence intensity profile measured using PIV in figure~\ref{fig:stats}(b). 
The streamwise turbulence intensity measured using PIV appears to be consistently higher in magnitude than the hot-wire result which can be attributed to the relatively finer spatial resolution of the planar PIV and/or higher measurement noise. 

\begin{figure}
  \begin{center}
    \includegraphics[width=0.8\textwidth]{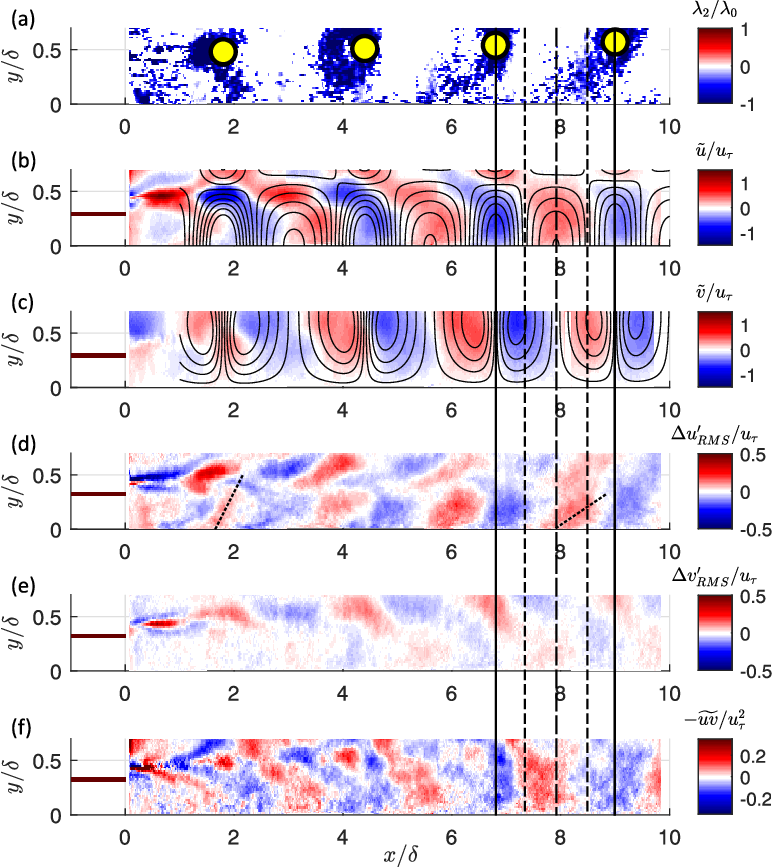}
    \caption{(a) $\lambda_{2}$ \eqref{eq:L2} (b) Streamwise component of modal velocity. (c) Wall-normal component of modal velocity. (d) Streamwise component of residual turbulence. (e) Wall-normal component of residual turbulence. (f) Phase-averaged Reynolds stress. $H=0.3\delta,\ f_{p}=80$ Hz.}
    \label{fig:maps}
  \end{center}
\end{figure}

To identify the locations of ALSSA-induced vortical structures, the $\lambda_{2}$-criterion defined by \citet{jeong_identification_1995} was used:
\begin{equation}
    \lambda_{2} = \frac{\partial \tilde{u}}{\partial x}^{2} + 
    \frac{\partial \tilde{v}}{\partial x} \frac{\partial \tilde{u}}{\partial y}.
    \label{eq:L2}
\end{equation}
\noindent These results are presented in figure~\ref{fig:maps}(a). 
The map of $\lambda_{2}$ has been normalized by the maximum absolute value (called $\lambda_{0}$) such that a value of $\lambda_{2}/\lambda_{0}=-1$ would indicate a clockwise rotating spanwise vortical structure in the flow. 
From this map of $\lambda_{2}$ it is now possible to identify the instantaneous location of the synthetic LSS in the outer region. 
The approximate locations of synthetic spanwise vortices are indicated with yellow circles in figure~\ref{fig:maps}(a) (corresponding to the regions of $\lambda_{2}/\lambda_{0}\approx-1$) which indicate a streamwise wavelength of approximately $\lambda_{x}=2\delta$ for the synthetic LSS. 
A single phase of the actuation cycle is shown in figure~\ref{fig:maps} which corresponds to the plasma being turned on ($\phi=0$). 

Spatially resolved maps of the streamwise and wall-normal components of modal velocity obtained via planar PIV are also presented in figure~\ref{fig:maps}(b,c). 
Here the physical location and dimension of the actuator plate is shown by the dark horizontal bar on the left-hand side of each map. 
Above the actuator there are distinct and periodic fluctuations in the streamwise modal velocity associated with the plasma jet which is generated on the top side of the actuator, figure~\ref{fig:maps}(b). 
There are also fluctuations, of comparable magnitude to the friction velocity, in the region between the actuator and the wall, with the highest amplitude primarily around the log-linear region (nominally $y^{+}=100,\ y/\delta=0.15$), that occur slightly downstream of the same-signed fluctuations above the actuator. 
The induced variations in the streamwise modal velocity below the actuator also extend nearly to the wall, though the amplitude of the variations begins to diminish significantly as the wall is approached. 
Similarly there are fluctuations in the wall-normal modal velocity, also comparable in magnitude to the friction velocity, across the entire wall-normal measurement extent, figure~\ref{fig:maps}(c). 
However, for the wall-normal modal velocity, the fluctuations above and below the actuator location are in phase, especially at the most downstream locations. 
These results are indicative of the distinct spanwise vortices induced by the plasma forcing and consistent with previous experimental single point hot-wire measurements \citep{lozier_experimental_2023} and results for a numerically simulated actuated boundary layer \citep{liu_experimental_2002}. 

These streamwise and wall-normal modal velocity fields were also modeled using Biot-Savart induction considering point vortex sources (and their corresponding images) which are representative of these synthetic LSS (i.e., the yellow circles in figure~\ref{fig:maps}a). 
Contours of the velocity fields generated from the Biot-Savart induction model are overlaid in figure~\ref{fig:maps}(b,c), with contour levels of ($-1.5:0.3:1.5$). 
With proper adjustment of the circulation and diffusion for each point vortex, the reconstructed modal velocities match in not only their spatial distribution, but also amplitude starting from around $x=4\delta$. 
The fluctuations in modal velocity closer to the actuator are not as well explained through this Biot-Savart model, as the induced velocity fluctuations immediately downstream of the actuator trailing edge, $x=0-2\delta$, may be influenced by secondary effects due to the local flow conditions (i.e., induced pressure gradients and/or circulation) around the actuator plate during plasma forcing. 
However, the agreement of the measured phase-locked velocities with this Biot-Savart model and the spatial input-output model of \citet{liu_spatial_2022} demonstrate that the superimposition of significant large-scale motions from the outer-region onto the log-linear and near-wall regions can be modeled through simple linear processes. 

\begin{figure}
\centering
    \includegraphics[width=0.85\textwidth]{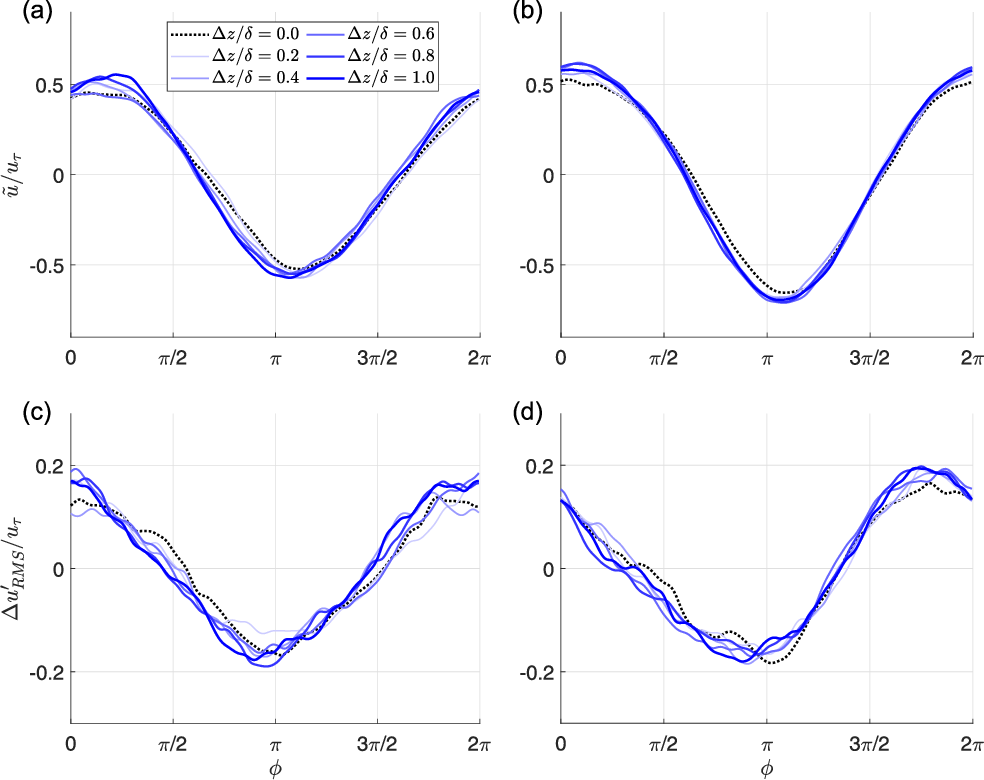}
    \caption{Phase dependent variations in (a,b) modal velocity and (c,d) residual turbulence at (a,c) $y^{+}=25$ and (b,d) $y^{+}=100$ across multiple spanwise locations. $H^{+}=200,\ x=8\delta,\ f_{P}=80$ Hz.}
    \label{fig:modal}
\end{figure}

It is also noteworthy that the streamwise and wall-normal modal velocity measured in this experiment, figure~\ref{fig:maps}(b,c), are similar in both amplitude and spatial distribution to the same quantities measured in experiments where large-scale motions were introduced via dynamic wall roughness \citep{huynh_characterization_2020}. 
The similarity in results demonstrate that the actuator design used here can achieve effective large-scale modification of the near-wall velocity without being physically located at, or near, the wall. 
Additionally, the streamwise modal velocity measured using the spanwise offset hot-wires at two representative wall-normal locations near the downstream end of the PIV domain are shown in figure~\ref{fig:modal}(a,b). 
Over separation distances, $\Delta z$, of up to one boundary layer thickness, the streamwise modal velocities measured in the near-wall and log-linear regions are identical (in amplitude and phase; within experimental uncertainty) to the measurement at the mid-span (which also agrees well with the mid-span PIV measurement). 
This confirms the extent of spanwise uniform flow which is induced by the actuator and indicates that the results from figure 6 are representative of the flow over a significant spanwise distance, not just the mid-span. 

In addition to the modal velocity, the components of residual turbulence are presented in figure~\ref{fig:maps}(d,e). 
In the streamwise residual turbulence, figure~\ref{fig:maps}(d), there is a visual distinction between fluctuations around the actuator height (a signature of the synthetic LSS themselves), and fluctuations in the near-wall region (the TBL response i.e., modulated turbulence intensity). 
This is especially clear just downstream of the actuator, $x=4\delta$, where two spatially distinct regions of positive residual turbulence fluctuation can be seen. 
Above the actuator height, $y/\delta=0.5$, the residual turbulence fluctuations associated with the synthetic LSS decay in the downstream direction. 
In contrast, the residual turbulence fluctuations below the actuator, which represent the TBL response to the imposed LSS, are growing in amplitude until approximately $x=7\delta$ downstream of the actuator where they also begin to decay. 
The maximum amplitude of residual turbulence below the actuator, across all streamwise locations, is consistently around the log-linear region of the TBL, $y^{+}\approx100,\ y/\delta\approx0.15$, and on the order of the friction velocity. 
This distinct region of modulated turbulence also changes shape, or orientation, moving downstream as shown by the dashed lines in figure~\ref{fig:maps}(d) which correspond to an inclination from the wall of $45^\circ$ (left) and $20^\circ$ (right). 
In previous experiments it was determined that the inclination of this region of modulated turbulence was similar to the typical inclination of attached eddies which can vary across the TBL between $15-45^\circ$ \citep{adrian_vortex_2000}, which is now confirmed here with spatially resolved measurements. 
This shape is also similar to, and aligns with, the vorticity pattern extending to the wall as shown in figure~\ref{fig:maps}(a). 
At the farthest downstream locations, starting around $x=8\delta$, the two previously distinct regions of residual turbulence fluctuations begin to blend or interact. 
These behaviors demonstrate an evolution or adjustment of the TBL from its canonical state in response to the presence of the synthetic LSS. 
Similar to the conclusions about the modal velocity, fluctuations in the residual turbulence near the wall immediately downstream of the actuator, should not be attributed directly to the synthetic LSS which are not yet well-defined (i.e., not fully developed). 

The streamwise component of the residual turbulence was also measured using the spanwise offset hot-wires as shown in figure~\ref{fig:modal}(c,d). 
Over separation distances of up to one boundary layer thickness, the streamwise residual turbulence also remains identical to the measurement at the mid-span of the tunnel. 
This again confirms the significant extent of spanwise uniform flow which is produced by the actuator, and shows that the TBL response is also spanwise uniform across this extent. 

For the wall-normal component of the residual turbulence, shown in figure~\ref{fig:maps}(e), significant fluctuations are only observed in the region above the actuator. 
Further, these fluctuations are of the same magnitude as the streamwise component in this region, and therefore also considered a direct signature of the synthetic LSS themselves. 
Critically, the absence of similar fluctuation signatures below the plate confirms that these synthetic vortical structures remain confined to the outer region and do not physically encroach into the near-wall or log-linear regions. 
This also indicates that the modulation of the streamwise residual turbulence observed below the actuator plate in figure~\ref{fig:maps}(d) is not an artifact of the synthetic LSS themselves, but induced through some interaction. 
The map of the phase-dependent Reynolds stress is shown in figure~\ref{fig:maps}(f). 
There are fluctuations in the Reynolds stress across the entire measurement region, and the fluctuations have a nearly constant amplitude across spatial locations. 
It is noted that in the log-linear and near-wall regions, positive peaks peaks in $-\widetilde{uv}/u^2_{\tau}$ lead peaks in residual turbulence, and by a downstream location of $x=7\delta$ the fluctuations have become somewhat organized in the wall-normal direction with a shape similar to the wall-normal modal velocity. 
It is also noted that the above PIV measurements, of all turbulence quantities, are consistent with the results presented in previous studies for single point hot-wires \citep{lozier_response_2024}. 

\begin{figure}
  \begin{center}
    \includegraphics[width=1\textwidth]{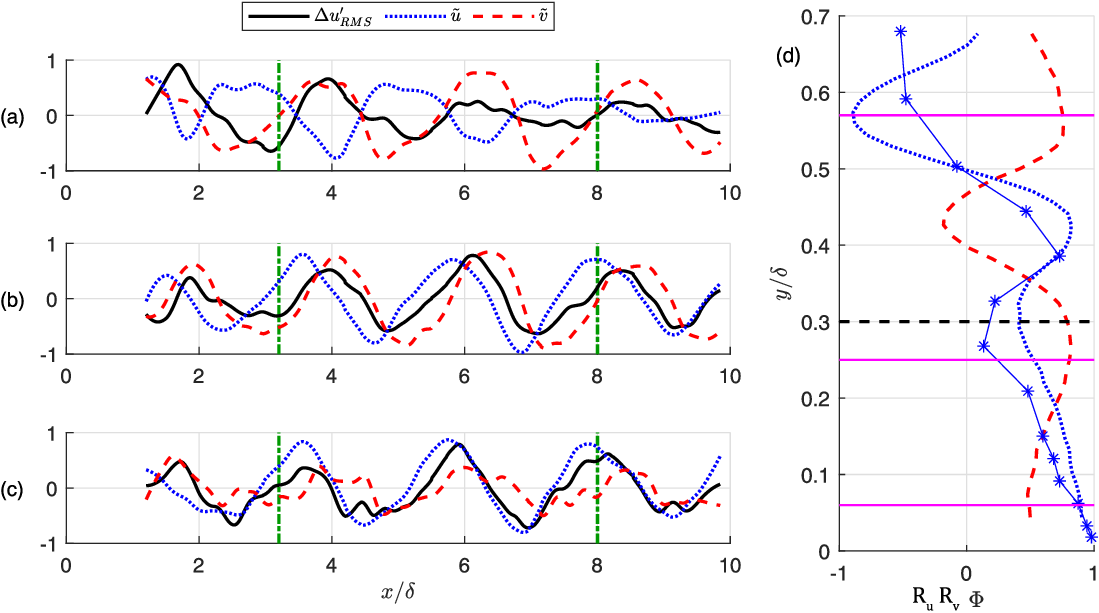}
    \caption{
    Normalized traces of streamwise modal velocity, wall-normal modal velocity and streamwise residual turbulence at (a) $y/\delta=0.57$ ($y^{+}=400$) (b) $y/\delta=0.24$ ($y^{+}=160$) and (c) $y/\delta=0.06$ ($y^{+}=40$). (d) Normalized correlation between residual turbulence and modal velocities ($R_{u}$ $\rightarrow$ blue dotted line, $R_{v}$ $\rightarrow$ red dashed line). Vertical green dash-dotted lines in (a,b,c) indicate approximate bounds for correlation region. Blue asterisks are $\Phi$ modulation coefficient from \cite{lozier_response_2024}. Solid magenta lines correspond to locations of traces in (a,b,c). Actuator location marked by horizontal dashed line. $H=0.3\delta,\ f_{p}=80$ Hz.}
    \label{fig:xcorr}
  \end{center}
\end{figure}

To further investigate the relationship between fluctuations in the streamwise residual turbulence and both components of modal velocity, streamwise traces of each quantity have been extracted at three representative wall-normal locations in figure~\ref{fig:xcorr}(a,b,c). 
The normalized spatial correlations between each component of the modal velocity ($\tilde{u},\ \tilde{v}$) and the change in residual turbulence ($\Delta u'_{RMS}$) across all wall-normal locations, 
\begin{equation}
    R_{u_{i}}(y) = \frac{\langle \tilde{u_{i}}(x,y,\phi)\Delta u'_{RMS}(x,y,\phi)\rangle_{x}}
    {\sqrt{\langle \tilde{u_{i}}(x,y,\phi)^{2}\rangle_{x}}
    \sqrt{\langle \Delta u'_{RMS}(x,y,\phi)^{2}\rangle_{x}}},
    \label{eq:Ry}
\end{equation}
\noindent are shown in figure~\ref{fig:xcorr}(d) as a complement to the velocity traces. 
At the location closest to the wall ($y^+=40$), shown in figure~\ref{fig:xcorr}(c), the periodic changes in residual turbulence are in phase with, but slightly lead, changes in the streamwise modal velocity. 
A similar phenomenon has also been reported for classical amplitude modulation effects measured in canonical high-Reynolds number TBLs \citep{mathis_large-scale_2009}. 
This is also reflected in figure~\ref{fig:xcorr}(d) where there is a positive correlation between both modal velocity components and the residual turbulence near the wall, indicating that they are in phase. 
The spatial correlation between the streamwise modal velocity and residual turbulence ($R_{u}$ from \eqref{eq:Ry}) is also analogous to the modulation coefficient denoted as $\Phi$ \eqref{eq:Phi}, which is a phase-based correlation that has been used for single point hot-wire measurements in \cite{lozier_response_2024}, 
\begin{equation}
    \Phi(y)=\frac{\langle\tilde{u}(x,y,\phi)
    \Delta u'_{RMS}(x,y,\phi)\rangle_{\phi}}
    {\sqrt{\langle\tilde{u}(x,y,\phi)^{2}\rangle_{\phi}}\sqrt{\langle\Delta u'_{RMS}(x,y,\phi)^{2}\rangle_{\phi}}}.
    \label{eq:Phi}
\end{equation}
\noindent Agreement between $\Phi$ (blue asterisks) and $R_{u}$ (blue dotted line) across all wall-normal locations can be seen in figure~\ref{fig:xcorr}(d) highlighting that these results do not depend on the specific measurement technique or correlation method (i.e., spatial versus temporal correlations). 
It is also noted that in previous experiments the skewness magnitude for the actuated TBL was higher than the baseline canonical case in the near-wall and log-linear regions supporting, broadly, an increase in inter-scale coupling under plasma forcing, \cite{Midya_2025}. 
The phase reversal reported for high-Reynolds number canonical TBLs \citep{mathis_large-scale_2009} is not observed across the log-linear region here, figure~\ref{fig:xcorr}(d), however there is a peak positive correlation between the residual turbulence and the wall-normal modal velocity, $R_{v}$ in this region (which leads in phase, figure~\ref{fig:xcorr}b). 
This suggests that both the large-scale wall-normal and streamwise motions may contribute to the observed modulation of streamwise turbulence intensity, depending on the wall-normal location. 
There is an eventual phase reversal \citep{mathis_large-scale_2009} and negative correlation between the streamwise modal velocity and residual turbulence, but only well above the actuator location as shown in figure~\ref{fig:xcorr}(a,d). 

\begin{figure}
  \begin{center}
    \includegraphics[width=0.8\textwidth]{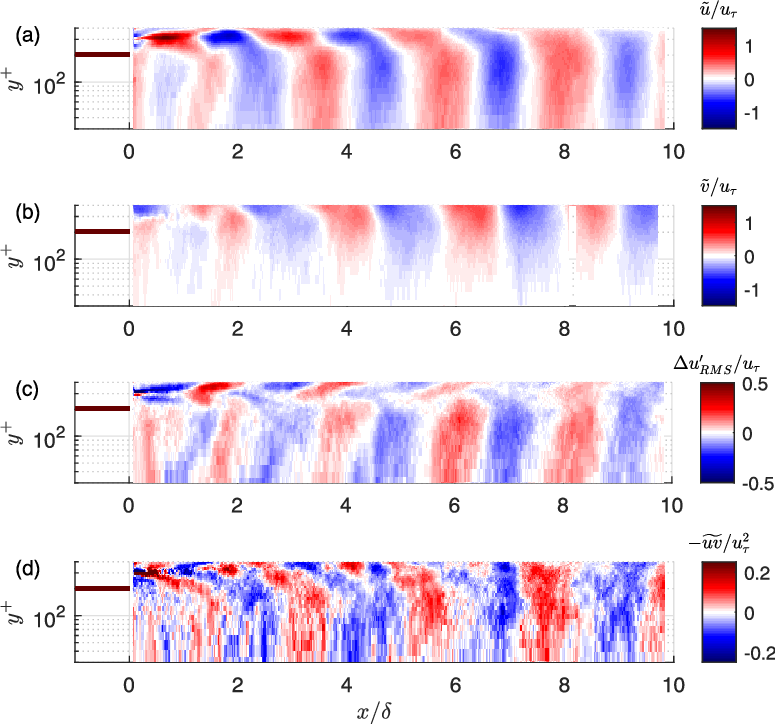}
    \caption{(a) Streamwise component of modal velocity. (b) Wall-normal component of modal velocity. (c) Streamwise component of residual turbulence. (d) Phase-locked Reynolds stress. $H^{+}=200,\ f_{p}=80$ Hz.}
    \label{fig:inner}
  \end{center}
\end{figure}

The streamwise and wall-normal components of the modal velocity are re-plotted in figure~\ref{fig:inner}(a,b) where the wall-normal, y-coordinate, has been converted to inner-variable scaling and the y-axis scale is logarithmically scaled in order to better visualize the region of interest below the actuator. 
The minimum wall-normal distance shown in the maps in figure~\ref{fig:inner} is $y^{+}=30$. 
The map of streamwise residual turbulence downstream of the actuator is also re-plotted in figure~\ref{fig:inner}(c) focusing attention on the region which is being modulated by the synthetic LSS i.e., the near-wall TBL response to the imposed large-scale perturbation. 
The distinct shape or inclination of the modulated turbulence is apparent where the higher amplitude fluctuations occur near the log-linear region ($y^{+}=100$) and also occur farther downstream than the related fluctuations closer to the wall. 
It can also be seen that for a single pairing of negative and positive fluctuations in the residual turbulence (between $x=6-7\delta$ for instance), the transition from negative to positive amplitude in the upstream direction occurs over a shorter streamwise distance than the opposite transition from a positive to negative amplitude. 
This suggests that some nonlinear process may be contributing to the modulation in this region, otherwise the fluctuations in residual turbulence would be expected to follow a more evenly periodic shape like that seen in the wall-normal modal velocity, for instance. 
The positive correlation between the residual turbulence and wall-normal modal velocity away from the wall can also be observed suggesting that a wall-normal transport mechanism could also link the large-scale motions and the modulated turbulence in the log-linear region. 

Phase-locked variations in the Reynolds stress have been re-plotted in figure~\ref{fig:inner}(d). 
The strongest changes in the Reynolds stress are again concentrated around the log-linear region and have a similar asymmetrical periodicity as the streamwise residual turbulence. 
In this case the variations in the Reynolds stress appear to occur slightly upstream of the changes in the residual turbulence, or it could be said that peak Reynolds stress lags peak streamwise residual turbulence, but are in phase with changes in the streamwise modal velocity near the wall. 
A changing mean streamwise velocity gradient, due to the streamwise modal velocity fluctuations, and changes in Reynolds stress amplitude could be indications of a change in the local turbulence production near the wall, though the current measurements are not spatially resolved enough to directly measure this. 

\begin{figure}
  \begin{center}
    \includegraphics[width=0.8\textwidth]{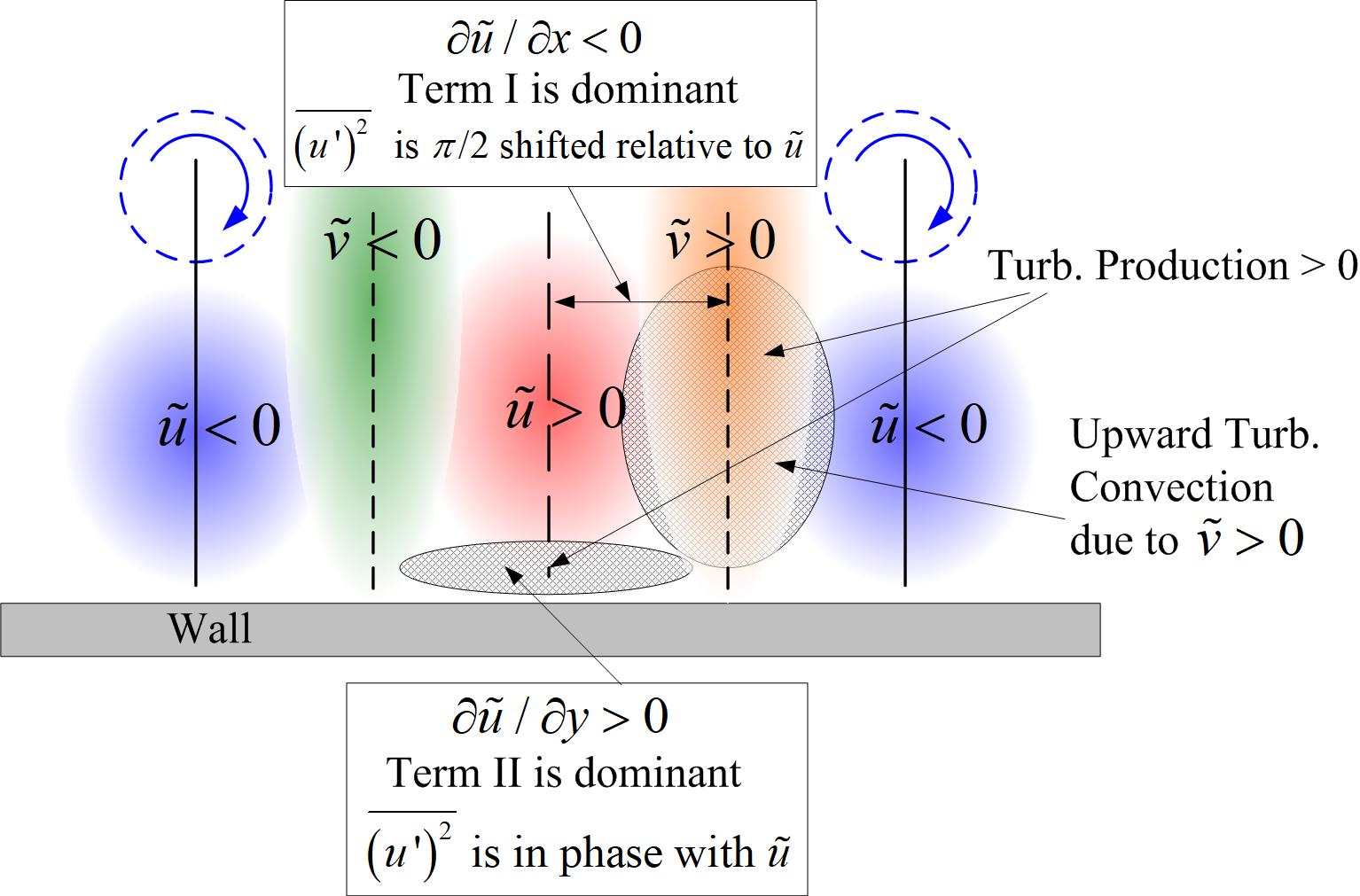}
    \caption{Schematic of the energy production due to the vortical structures induced by ALSSA. Vortical structures move from left to right.}
    \label{fig:production}
  \end{center}
\end{figure}

In \cite{Reynolds_Hussain_1972}, a triple Reynolds decomposition was implemented to derive the governing dynamic equations describing the interaction between the mean flow, organized motions and turbulence. 
One term of interest for this research is the contribution to the turbulence production from the modal velocity components due to the synthetic vortical structures, 
\begin{equation}
\frac{D}{Dt} \left (\overline{\frac{u'_i u'_i}{2} } \right ) =...-\overline{\langle u'_i u'_j \rangle  \frac{\partial \tilde{u_i}}{\partial {x_j}}}...,
\end{equation}
\noindent where the overbar denotes time-averaging, and the angled brackets denote the phase-averaging.
For the case of the two-dimensional modal field, the contribution to the streamwise component fluctuation by the turbulent production becomes,
\begin{equation}
\frac{D}{Dt} \left ( \frac{\overline{u'^2}}{2} \right ) =...- \underset {\text {Term I} } {\overline {\langle u'^2 \rangle\frac{\partial \tilde{u}}{\partial x}} }- \underset {\text {Term II} } { \overline { \langle u'v' \rangle\frac{\partial \tilde{u}}{\partial y} } }...
\label{eq:Production}
\end{equation}
\noindent A schematic of the contribution to the streamwise component of the modal velocity due to the periodic vortical structure is presented in figure~\ref{fig:production}. 
The modal velocity is negative immediately below the vortical structure via the Biot-Savart mechanism (indicated as a blue regions in figure~\ref{fig:production}), and is positive half-way between, or $\pi$-out-of-phase with, the adjacent structures (indicated as a red region in figure~\ref{fig:production}). 
This schematic is consistent with the experimental results presented in figure~\ref{fig:maps}. 

The dilatation-related Term I in \eqref{eq:Production} is positive in the region where ${\partial \tilde{u}}/{\partial x} <0$. 
From figure~\ref{fig:production} it follows that the positive turbulence production related to the Term I is $\pi/2$ upstream of the vortical structure. 
As a consequence, the residual turbulence is $\pi/2$ out of phase with the modal u-component in the region below the actuator, which is, in this case, in the log-linear region. 
This is consistent with smaller values of $R_u$ (and the $\Phi$-modulation coefficient), which describes the phase difference between the residual turbulence and the streamwise modal velocity, in the log-linear region in figure~\ref{fig:xcorr}(d). 

The shear-related Term II in \eqref{eq:Production} is positive near the wall where ${\partial \tilde{u}}/{\partial y} >0$, since $\langle u'v' \rangle<0$. 
As at the wall ${\tilde{u}(0)=0}$, in the near-wall region $\partial \tilde{u}/{\partial y} \sim \tilde{u}$. 
It implies that the residual turbulence should be in phase with the streamwise component of the modal velocity near the wall, which is consistent with close-to-one values of the $\Phi$-modulation coefficient in the near-wall region in figure~\ref{fig:xcorr}(d). 
Finally, in addition to being produced, the turbulence is also transported upward in the region where $\tilde{v}>0$, as indicated in figure~\ref{fig:production}, extending the region of residual turbulence in the wall-normal direction, even above the location of ALSSA, see figure \ref{fig:maps}(d) or figure \ref{fig:inner}(c). 

These combined results then create a physical description of the relationship/interaction between the induced large-scale vortical motions in the outer-region and modulated turbulence intensities in the near-wall and log-linear regions. 
Large-scale motions induced by these vortices are superimposed onto the near-wall region through a linear process consistent with Biot-Savart induction. 
These superimposed large-scale motions, and their distinct spatial distribution, then lead to two regions of phase-dependent turbulence production, through different mechanisms. 
The induced large-scale wall-normal fluctuations are further responsible for the transport turbulence away from the wall, especially within the log-linear region. 
These combined turbulence production and transport mechanisms lead to the observed modulation of turbulence intensity in the near-wall and log-linear regions, which are in phase with the synthetic LSS. 
While the large-scale linear superposition and shear-related production mechanisms are generally consistent with classical amplitude modulation effects, it is unclear if comparable counterparts to the wall-normal transport and dilatation-related production mechanisms exist for canonical LSS, as they are not typically associated with classical amplitude modulation effects. 
These result also motivate additional experiments, with greater spatial resolution, in order to measure these mechanisms of turbulence production and/or transport explicitly/independently. 

\begin{figure}
  \begin{center}
    \includegraphics[width=1\textwidth]{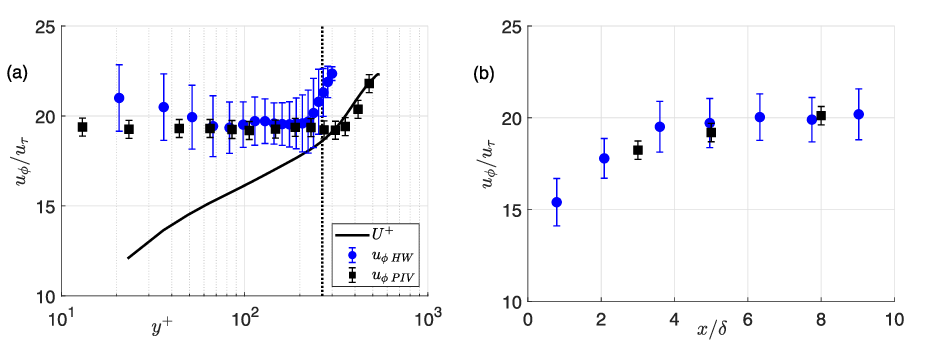}
    \caption{Profiles of mean velocity (solid black line) and phase-speed using a hot-wire from \citet{lozier_response_2024} (black squares) and PIV (blue circles), with $90\%$ confidence intervals. $x=5\delta$. (b) Streamwise development of phase-speed measured at $y^{+}=100$. $H^{+}=200,\ f_{p}=80$ Hz.}
    \label{fig:phasespeed}
  \end{center}
\end{figure}

We now investigate how these large-scale motions interact with, and/or change the characteristics of, the canonical near-wall structure/cycle.
First, the phase-speed of large-scale fluctuations was estimated by computing a streamwise spatial cross-correlation, $R_{x}$, of the streamwise component of the modal velocity,
\begin{equation}
    R_{x}(\Delta x,y,\phi,\Delta \phi) = \frac{\langle \tilde{u}(x,y,\phi) 
    \tilde{u}(x+\Delta x,y,\phi+\Delta \phi)\rangle_{x}}
    {\langle \sqrt{\tilde{u}(x,y,\phi)^{2}\rangle_{x}}
    \langle \sqrt{\tilde{u}(x,y,\phi+\Delta \phi)^{2}\rangle_{x}}}.
    \label{eq:Rx}
\end{equation}
\noindent A maximum in this streamwise spatial cross-correlation ($R_{x,max}$) represents the streamwise distance ($\Delta x$) a large-scale fluctuation travels during the small time interval between two PIV measurements ($\Delta \phi T_{P}/2\pi$). 
As described in section~\ref{sec:setup}, the PIV measurements resulted in five discrete phases (i.e., $\phi=n\pi/5$ with $n=0,2,4,6,8$). 
As such, the cross-correlation was repeated for each of these distinct phases, with a fixed $\Delta \phi=2\pi/5$, and then averaged to improve the cross-correlation estimate in each PIV measurement region. 
The streamwise displacement which results in a maximum in the cross-correlation, over a given time interval, can be used to determine the average phase-speed of large-scale fluctuations over each measurement region,
\begin{equation}
    u_{\phi PIV}(y)=\frac{\Delta x|_{R_{x}(\Delta x,y,\Delta \phi)=R_{x,max}}}{\Delta \phi T_{P}/2\pi}.
    \label{eq:phasespeed_PIV}
\end{equation}
\noindent The results of this phase-speed analysis are shown in figure~\ref{fig:phasespeed} as blue circles. 
There is good agreement in the estimated phase-speed between the single point hot-wire measurements and PIV across all streamwise locations, especially within the log-linear region of the boundary layer as demonstrated in figure~\ref{fig:phasespeed}(b). 
In figure~\ref{fig:phasespeed}(a), above the actuator height, the phase-speed of large-scale phase-locked fluctuations appear to be converging with the local mean velocity. 
Below the actuator height the phase-speed appears to be nearly constant with a magnitude that is similar to the mean velocity at the location of the synthetic LSS (i.e., the actuator height). 
This observation indicates that fluctuations below the actuator are governed by the convection of the synthetic large-scale structures produced in the outer region  rather than following the local dynamics (i.e., the local mean velocity), resulting in a top-down traveling wave type effect. 
In figure~\ref{fig:phasespeed}(b) the phase-speed around the log-linear region, $y^{+}=100$, appears to be reaching an asymptote by the farthest downstream locations. 
This phase-speed asymptote characterizes the time scale of the boundary layer's adjustment to the presence of a synthetic large-scale structure, and corresponds well to the increasing magnitude of residual turbulence amplitudes observed in figure~\ref{fig:modal}(d). 
This observation also motivates additional measurements over a larger downstream fetch, as the response of the TBL to the synthetic LSS evolves significantly over the present measurement domain. 
Additionally, the phase-speed can be calculated using any phase-locked quantity, but in previous studies they all showed the same trend \citep{lozier_response_2024}, so the streamwise modal velocity was chosen here for simplicity. 

\begin{figure}
\centering
    \includegraphics[width=0.85\textwidth]{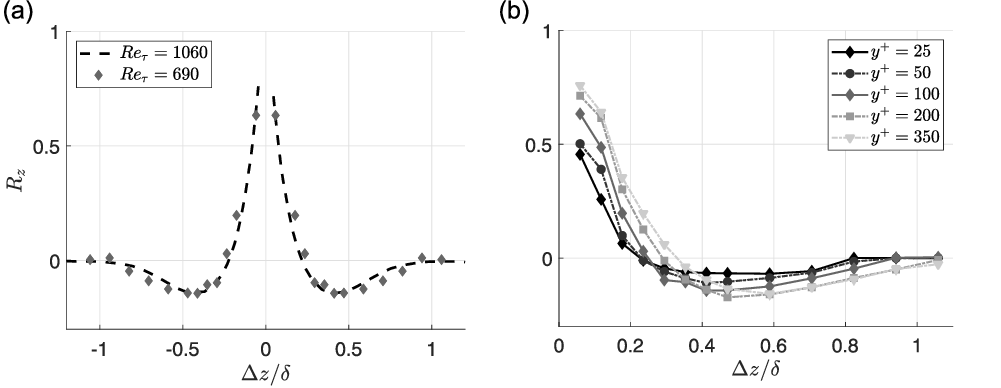}
    \caption{Spanwise profiles of the streamwise velocity correlation for the canonical TBL at (a) $y^+=100$ overlaid with data from \cite{ganapathisubramani_investigation_2005} ($Re_{\tau}=1060,\ y^{+}=92$) and (b) a range of wall-normal locations with lighter colors indicating greater distance from the wall. $x=8\delta$.}
    \label{fig:canon_corr}
\end{figure}

To investigate what effect the synthetic LSS has on the spanwise organization of near-wall structures and the dynamics of near-wall events, velocity correlations were computed using two spanwise offset hot-wires. 
To quantify the spanwise correlation of the streamwise velocity between the two hot-wire probes, a standard two-point normalized correlation coefficient was computed,
\begin{equation}
    R_{z}(x,y,\Delta z) = \frac{\overline{u(x,y,z,t)u(x,y,z+\Delta z,t)}}{\sqrt{\overline{u(x,y,z,t)}^{2}}\sqrt{\overline{u(x,y,z+\Delta z,t)}^{2}}}.
    \label{eq:spancorr}
\end{equation}
\noindent An example spanwise correlation coefficient profile for the canonical TBL (no actuator) obtained at $x/\delta = 8$ is shown in figure~\ref{fig:canon_corr}(a). 
In the log-linear region ($y^{+}=100$) the first crossing from positive to negative correlation is around $\Delta z/\delta=0.25$ which is consistent with the typical spanwise scale for $\delta$-scaled vortical structures \citep{tomkins_spanwise_2003, ganapathisubramani_investigation_2005}. 
The data is plotted along with published results for a canonical boundary layer with low-$Re_{\tau}$ from \citet{ganapathisubramani_investigation_2005}, at a similar wall-normal location, and good agreement between the profiles is observed. 
The spanwise correlation coefficient profiles for the baseline canonical TBL across a range of wall-normal locations are shown in figure~\ref{fig:canon_corr}(b). 
Looking at the first zero-crossing of each profile, there is a clear growth in spanwise coherence, which is dependent on wall-normal location, from $y^{+}=25$ up to $y^{+}=350$ ($y/\delta=0.5$) consistent with previous studies \citep{tomkins_spanwise_2003}. 
Closer to the wall the zero-crossing is approaching the reported value of $\Delta z^{+}=100$ ($\Delta z/\delta=0.15$) which corresponds to the distribution of canonical near-wall structures \citep{antonia_spanwise_1990}. 

\begin{figure}
\centering
    \includegraphics[width=0.85\textwidth]{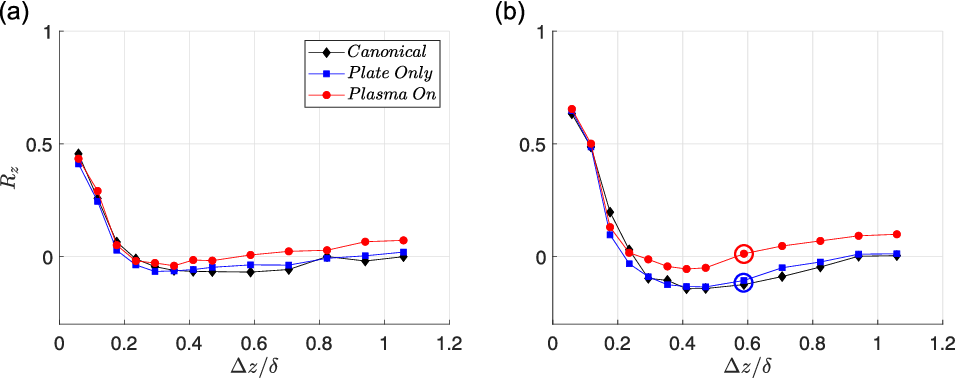}
    \caption{Spanwise profiles of streamwise velocity correlation for canonical, plate only, and plasma on cases at (a) $y^{+}=25$ and (b) $y^{+}=100$. Circled points correspond to horizontal lines in figure~\ref{fig:phase_corr}. $H^{+}=200,\ x=8\delta,\ f_{P}=80$ Hz.}
    \label{fig:compare_corr}
\end{figure}

Spanwise correlations for the canonical, plate only, and plasma on cases in the near-wall and log-linear regions are compared in figure~\ref{fig:compare_corr}. 
At both wall-normal locations, the addition of the actuator plate alone does not result in a modification of the level of spanwise correlation as compared to the canonical case. 
However, when the plasma forcing is active, there is a positive shift in the spanwise correlation, even at the largest separation distances. 
In both locations, but specifically in the log-linear region, the first zero crossing of the profile does not change location indicating that there is no net effect of the actuator plate or plasma forcing on the spanwise organization of small-scale vortical structures. 

\begin{figure}
\centering
    \includegraphics[width=0.5\textwidth]{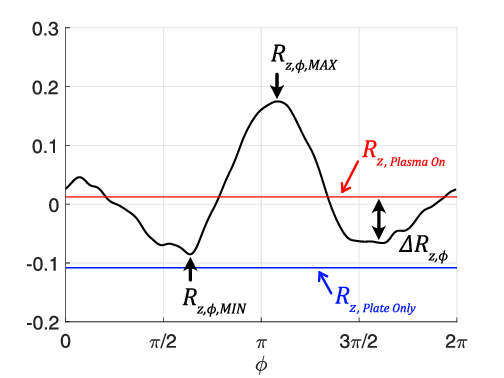}
    \caption{Phase dependent variation in spanwise correlation with $\Delta z=0.6\delta$ for plasma on case at $y^{+}=100$. Horizontal lines correspond to circled points in figure~\ref{fig:compare_corr}(b). $H^{+}=200,\ x=8\delta,\ f_{P}=80$ Hz.}
    \label{fig:phase_corr}
\end{figure}

However, this correlation is an average measure, and in the case of active plasma forcing it is useful to decompose the velocity time series which has been phase-locked to the plasma forcing. 
Using the phase-locked decomposition method shown earlier, a phase dependent two-point correlation coefficient was also computed, 
\begin{equation}
    R_{z,\phi}(y,\Delta z,\phi) = \frac{\langle u(y,z,\phi,n)u(y,z+\Delta z,\phi,n)\rangle_{n}}{\sqrt{\langle u(y,z,\phi,n)\rangle_{n}^{2}}\sqrt{\langle u(y,z+\Delta z,\phi,n)\rangle_{n}^{2}}}.
    \label{eq:spancorrp}
\end{equation}
\noindent As a final step, similar to the residual turbulence, the phase average of this phase-dependent correlation was removed in order to analyze phase-dependent changes in the spanwise correlation, later labeled as $\Delta R_{z,\phi}$ (figure~\ref{fig:phase_corr}). 
A phase-locked measurement of the spanwise correlation in the log-linear region ($y^{+}=100$) for a single spanwise separation distance is shown in figure~\ref{fig:phase_corr}. 
The average spanwise correlations for the plasma on and plate only cases are shown as horizontal lines which correspond to the encircled points in figure~\ref{fig:compare_corr}(b). 
The peak in the phase dependent spanwise correlation ($R_{z,\phi,MAX}$), and minimum, ($R_{z,\phi,MIN}$), were identified as shown in figure~\ref{fig:phase_corr}. 
Comparison with figures~\ref{fig:modal}(b,d) reveals that the maximum in the phase-dependent spanwise correlation is aligned with negative fluctuations in both the streamwise modal velocity and residual turbulence.  
Additionally, the maximum spanwise correlation occurs just upstream of i.e., a phase lag with respect to, the synthetic LSS passing in the outer region. 

\begin{figure}
\centering
    \includegraphics[width=0.85\textwidth]{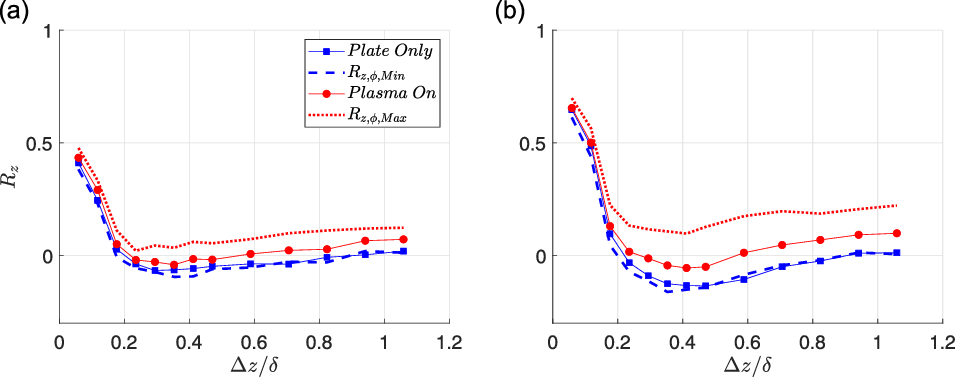}
    \caption{Spanwise profiles of streamwise velocity correlation for plate only and plasma on cases at (a) $y^{+}=25$ and (b) $y^{+}=100$. Maximum and minimum phase dependent correlation shown as dashed and dotted lines respectively. $H^{+}=200,\ x=8\delta,\ f_{P}=80$ Hz.}
    \label{fig:full_corr}
\end{figure}

The maximum and minimum amplitude of the spanwise correlation can then be computed across all spanwise separation distances and used to isolate effects relative to the passing of the synthetic LSS in the outer region. 
These results are shown in figure~\ref{fig:full_corr} for the streamwise location of $x=8\delta$ downstream of the actuator. 
Even sufficiently far downstream, the synthetic LSS still have a significant effect on the spanwise correlation as they pass through, demonstrated by the red dotted lines ($R_{z,\phi,MAX}$) in figure~\ref{fig:full_corr}.  
However, after the synthetic LSS moves slightly downstream, the correlation magnitudes relax back to canonical (or plate only) levels as shown by the dashed blue lines ($R_{z,\phi,MIN}$, which can also be seen in figure~\ref{fig:phase_corr}). 
These localized increases in the spanwise correlation (compared to the canonical/plate only magnitudes) over large spanwise separations, suggest that the spanwise-uniform, synthetic, large-scale motions induced by the actuator become dominant flow features of the near-wall and log-linear regions as the synthetic LSS convects past a streamwise location. 
Additionally, phases associated with dominant spanwise-uniform motions are also associated with a decrease in the residual turbulence, consistent with the model of \citet{Schoppa-Hussain_2002}. 
But this influence is also transient as the canonical spanwise correlation profile is recovered in between each passing. 
These combined phase-speed and spanwise correlation results suggest that locally there are changes to the near-wall structure dynamics as the induced spanwise-uniform large-scale motions travel through near-wall region; however there is no lasting modification of the near-wall structure as seen from the spanwise correlation recovery. 
This is further supported by \citet{Midya_2025} where both the frequency and intensity of near-wall bursting events were found to vary over the actuation cycle, with little net effect. 

% CONCLUSIONS =========================================
\section{Conclusions and Discussion}
\label{sec:disc}

An active plasma-based actuator was used to experimentally introduce periodic, spanwise-uniform, synthetic large-scale vortical structures into the outer region of a canonical turbulent boundary layer. 
The Reynolds number was kept sufficiently low so that energetic organized large-scale structures were not naturally present (as evidenced by premultiplied streamwise 1-D wavenumber spectra). 
In this manner, the top-down influence of these imposed, organized, outer region motions on the TBL could be effectively observed. 

Planar PIV and spanwise offset hot-wires were used to study the development of the synthetic LSS as well as the TBL’s response to this imposed large-scale perturbation. 
Spatially resolved PIV measurements showed that downstream of the actuator, spanwise vortices generated by the actuator induce coherent streamwise and wall-normal velocity fluctuations in the logarithmic and near-wall regions through a linear process consistent with Biot-Savart induction. 
These induced large-scale motions were also found to be directly correlated with local modulations of the small-scale turbulence intensity, similar to the classical amplitude modulation effect observed in higher Reynolds number canonical TBLs. 
Measurements indicate that the modulation of turbulence amplitudes here was associated with an imposed periodic variation in turbulence production, associated with both the wall-normal modal strain rate ($\partial\tilde{u}/\partial y$) and the dilatational modal strain rate ($\partial\tilde{u}/\partial x$), as well as wall-normal turbulence transport, associated with the wall-normal modal velocity component ($\tilde{v}$). 
While the superposition and shear-related production mechanisms are consistent with classical amplitude modulation effects, it is unclear if the wall-normal transport and/or dilatation-related production mechanisms are compatible with canonical LSS. 

The phase-speed of the induced large-scale motions was also found to be nearly constant below the actuator at a value of $u_{\phi}\approx0.84U_{\infty}$, which corresponds to the measured convection velocity of the synthetic LSS in the outer region. 
Measurements of the spatial coherence of streamwise velocity fluctuations showed that, in the presence of the synthetic LSS, the induced spanwise-uniform streamwise modal velocity fluctuations became a dominant feature of the flow, as evidenced by a significant positive shift in the spanwise spatial correlation, even at the largest probe separation distances. 
This was also shown to be locally correlated with phase-dependent modulation of turbulence intensity and the amplitude and frequency of bursting events \citep{Midya_2025}, collectively indicating a clear top-down influence of the synthetic LSS on the near-wall dynamics. 
However, in both the near-wall and log-linear regions, the first zero crossing of the correlation profile did not change in spanwise offset and the profile was found to relax back to canonical levels during phases of the actuation cycle corresponding to the synthetic LSS moving downstream from the measurement location indicating that there is no \emph{net} effect of the plasma forcing on the spanwise organization of smaller-scale vortical/streak structures. 
This is also reflected in the time-averaged streamwise turbulence intensity near the wall which does not change significantly under plasma forcing, for streamwise locations $x \lesssim 8\delta$. 
While these results collectively show a significant and measurable top-down influence of the outer region synthetic LSS on near-wall turbulence dynamics, as hypothesized, a lack of \emph{sustained alteration} to the near-wall structure suggests that it largely remains autonomous between observed intermittent interactions with the synthetic LSS. 

Additionally, these results/conclusions were found to be consistent with the previous work of \citet{liu_spatial_2022, lozier_response_2024} and the complementary study of \citet{Midya_2025} which further examined the skewness of the streamwise velocity fluctuations as a function of frequency (scale) in both the canonical and the actuated TBL under consideration here. 
The intimate connection between skewness and classical amplitude modulation effects in canonical TBLs was previously characterized by \citet{mathis_large-scale_2009, mathis_relationship_2011}. 
In these studies, a normalized amplitude modulation correlation coefficient, between low-pass filtered velocity signals in the outer region and the envelope of near-wall small-scale velocity fluctuations obtained via a Hilbert transform, was found to bear a very strong resemblance to the corresponding velocity skewness profile. 
Using bispectral methods \citep{Rosenblatt_1965, Elgar_1985}, the spectral distribution of skewness contributing scales for the plasma on and canonical boundary layers, presented in this work, were calculated and compared at selected wall-normal locations in \citet{Midya_2025}. 
It was found that in the logarithmic region ($y^{+}=60$) the spectral distributions of the skewness were distinctly different for the plasma on and canonical cases. 
However, in the near-wall region ($y^{+}=8$ and $15$), there remained a strong resemblance between the plasma on and canonical cases supporting the persistence of an autonomous structure/cycle in this near-wall region (consistent with the conclusions of the present study). 
\citet{Midya_2025} further showed that the skewness magnitude increases with Reynolds number across the near-wall and logarithmic regions of canonical turbulent boundary layers from different facilities spanning a full decade range of Reynolds numbers, including the canonical case investigated here. 
A similar increase in skewness magnitude was also observed for the plasma on case, as compared to the baseline canonical case, which can been attributed to the superimposition of large-scale motions in these regions \citep{Metzger_2001} and/or classical amplitude modulation effects \citep{mathis_large-scale_2009}. 
These skewness studies then confirm that while there is a clear top-down influence of the outer synthetic LSS by superimposing large-scale motions and modulating turbulence amplitudes/dynamics (i.e., frequency and amplitude of bursting event), the canonical autonomous near-wall structures/mechanisms remain dominant. 

These results also have implications regarding the design and optimization of future flow control techniques where top-down type actuation might be achievable and/or efficient. 
The present results indicate that transient changes in the near-wall turbulence amplitudes/dynamics can be achieved through the introduction of large-scale perturbations in the outer region, but there is a limitation in achieving sustained modification of near-wall structures/mechanisms, especially compared to actuation done at (or near) the wall \citep{marusic_energy-efficient_2021, Meyers_2025}. 
These results also motivate additional experiments to connect the processes of top-down interaction described here (i.e., for a uniform large-scale perturbation in the outer region) with the processes that lead to classical amplitude modulation effects in the near-wall region of canonical TBLs (i.e., from canonical LSS). 
This could involve the replication/modeling of additional characteristics of these canonical LSS like their inherently stochastic and three-dimensional distribution (i.e., meandering groups/packets of hairpin-like vortices) and/or testing higher-Reynolds number baseline TBLs which contain energetic large-scale motions and an established log-linear region.

% BACK MATTER =========================================

\backsection[Funding]{This research was funded by the Office of Naval Research, Grant: N00014-18-1-2534}

\backsection[Acknowledgements]{The authors would like to thank Dr. Igal Gluzman for his help in setting up and aligning the PIV system.}

\backsection[Declaration of interests]{The authors report no conflict of interest.}

\bibliographystyle{jfm}
\bibliography{references, additional_citations}

\end{document}